\documentclass[fleqn,twoside,onecolumn,nofootinbib,showkeys,12pt]{revtex4}

\usepackage{graphicx}

\usepackage{bm}

\begin{document}

\begin{center}

{\large \bf  Thermodynamically Anomalous Regions and  Possible New   Signals of  Mixed Phase Formation}

\vspace*{11mm}

{\bf K.A. Bugaev$^{1,*}$, A.I. Ivanytskyi$^1$, D.R. Oliinychenko$^{1,2}$,  V.V.~Sagun$^1$,  
I.N.~Mishustin$^{2, 3}$, D.H. Rischke$^4$, L.M. Satarov$^{2, 3}$ and  G.M.~Zinovjev$^1$}

\vspace*{5.5mm}

{\small

$^1${Bogolyubov Institute for Theoretical Physics, of the National Academy of Sciences of Ukraine, Metrologichna str. 14$^B$, Kiev 03680, Ukraine}\\

$^2${FIAS, Goethe-University,  Ruth-Moufang Str. 1, 60438 Frankfurt upon Main, Germany}\\

$^3${Kurchatov Institute,  Akademika Kurchatova Sqr., Moscow, 123182, Russia}\\

$^4${Institute for Theoretical Physics, Goethe-University,  Ruth-Moufang Str. 1, 60438 Frankfurt upon Main, Germany}\\
}

$^*${e-mail: bugaev@th.physik.uni-frankfurt.de}


\end{center}

{\bf Abstract.}{
Using an advanced version of the hadron resonance gas model we have found
indications for irregularities in data for hadrons produced in
relativistic heavy-ion collisions. These include
an abrupt change of the effective number of degrees of freedom,
a change of the slope of the ratio of lambda hyperons to protons 
at laboratory energies 8.6--11.6 AGeV, {as well as} highly correlated 
plateaus in the collision-energy dependence of the entropy per baryon, total pion 
number per baryon, and thermal pion number per baryon at 
laboratory energies 6.9--11.6 AGeV. Also, we observe a sharp 
peak in  the dimensionless  trace anomaly at a laboratory energy 
of 11.6 AGeV.
On the basis of the generalized shock-adiabat model we demonstrate  that these observations
give evidence for the  anomalous thermodynamic properties of the mixed phase 
at its boundary to the quark-gluon plasma.  
We argue that the trace-anomaly peak and the local minimum of the 
generalized specific volume observed at 
a laboratory energy of 11.6 AGeV 
provide a signal for the formation of a mixed phase
between the quark-gluon plasma  and the hadron phase.
This naturally explains the change of slope in the
energy dependence of the yield of
lambda hyperons per proton at a laboratory energy of 8.6 GeV.
}\\
{25.75.Nq}{ Quark deconfinement, quark-gluon plasma production, and phase transitions}\\
{25.75.-q}{ Relativistic heavy-ion collisions}\\
{25.75.Dw}{ Particle and resonance production}

\section{Introduction} 
During almost  thirty years of searching for the quark-gluon plasma (QGP) in heavy-ion collisions
 many {different} signals of its formation have been  
suggested.
{Among them  are some remarkable irregularities in 
excitation functions of experimental data,} known in the literature as the Kink \cite{Kink}, the 
Strangeness Horn \cite{Horn}, and the Step \cite{Step}. {Although these
irregularities are} often advocated as
signals of the onset of deconfinement \cite{Gazd_rev:10}, their relation to the 
QGP-hadron mixed phase is far from being clear. Therefore, 
additional and independent justification of these irregularities is required. 
This task is rather important in view of the planned heavy-ion collision experiments at 
JINR-NICA and GSI-FAIR.
Clearly, searching for other irregularities and signals of mixed-phase formation is no less 
significant, but until recently such efforts were not very  
successful, since they require a realistic model which is able to
accurately describe the existing experimental data and, thus, provide us with 
reliable information about the late stages of the 
heavy-ion collision process.

The recent extensions \cite{Oliinychenko:12,KAB:13,KAB:gs,NICAWP} of the hadron 
resonance gas model 
\cite{KABCleymans:93,KABCleymansFO,PBM:99,KABpbm:02,KABAndronic:05,KABAndronic:09}
provide us with the most successful description of available hadronic multiplicities measured
in heavy-ion collisions at AGS, SPS, and RHIC energies. 
The global values of $\chi^2/dof \simeq 1.16$  
and  $\chi^2/dof \simeq 1.06$ have been achieved, respectively, in 
Refs.\ \cite{KAB:13} and \cite{KAB:gs}  for 
111 independent multiplicity ratios measured at 14 values of  
the center-of-mass energy $\sqrt{s_{NN}} = $  2.7, 3.3, 3.8, 4.3, 4.9, 6.3, 7.6, 8.8, 9.2, 12, 17, 
62.4, 130, 200~GeV (for more details, see Ref.\ \cite{KAB:13,KAB:gs}).
This 
gives us confidence that the irregularities found in the narrow range of collision energies 
$\sqrt{s_{NN}} = 4.3-4.9$~GeV are not artifacts of the model and indeed reflect reality. 

However, the most difficult part of establishing 
a signal of mixed-phase formation is to 
formulate a convincing model of the heavy-ion collision process, which 
allows to connect the irregularity in 
the behavior of some observable with 
the occurrence of the QGP-hadron phase transition. Here we report such irregularities 
as an abrupt change of the effective number of degrees of freedom and a peak of the  
trace anomaly $\delta$, which are seen at $\sqrt{s_{NN}} = 4.9$~GeV,
and plateaus in the collision-energy dependence of the entropy per baryon, total pion 
number per baryon, and thermal pion number per baryon found  at 
laboratory energies 6.9--11.6 AGeV ($\sqrt{s_{NN}} = 3.8-4.9$~GeV). We observe all these 
irregularities at the stage of chemical freeze-out (FO) using the most successful 
version of the hadron 
resonance gas model  \cite{Oliinychenko:12,KAB:13,KAB:gs,NICAWP}. 
However, in order to relate them to the 
QGP-hadron phase transition we employ the model of the generalized shock adiabats 
for central nuclear collisions \cite{KAB:89a,KAB:90,KAB:91}, which 
predicted such plateaus a long time ago. 

Surprisingly, at chemical FO we also observe a second set of  plateaus at higher laboratory 
energies 30--44 GeV ($\sqrt{s_{NN}} = 7.6-9.2$~GeV), where the generalized shock-adiabat 
model is no longer reliable, but which is exactly the region  
where the onset of deconfinement is claimed to exist 
\cite{Kink,Horn,Step,Gazd_rev:10,RafLet}.  On the one hand, this fact demonstrates that
despite the large error bars 
the performed analysis of the entropy per baryon, total pion number per baryon, and thermal 
pion number per baryon extracted from the data at chemical FO is able to detect hidden 
correlations between the plateaus in these quantities. On the other hand, if the plateaus at 
laboratory energies 30--44 GeV are 
also generated by a mixed-phase instability, as it is the case for the set of 
plateaus found at lower collision energies, then this may 
provide evidence for a second phase transition. 

In addition, here we not only discuss the physical origin of novel signals of  
mixed-phase formation such as the minimum 
of the generalized specific volume and the sharp peak of 
the trace anomaly, we also relate them to the 
observed plato in the number of thermal pions  per baryon. 

This work is organized as follows. 
In the next section we report various irregularities existing at chemical FO. 
In order to provide a possible interpretation, in section 3 we 
present a short discussion of the generalized shock model and 
the possible origin of the above mentioned plateaus in the entropy per baryon, 
the total pion number per baryon, and 
the thermal pion number per baryon.  
We also present the details of extracting the parameters
of the plateaus from the data and from the chemical FO quantities. 
Section 4 is devoted to a justification of new signals of  
QGP-hadron mixed-phase formation, while our conclusions are summarized in section 5. 

\begin{figure}[t]
\centerline{\includegraphics[width=84.mm]{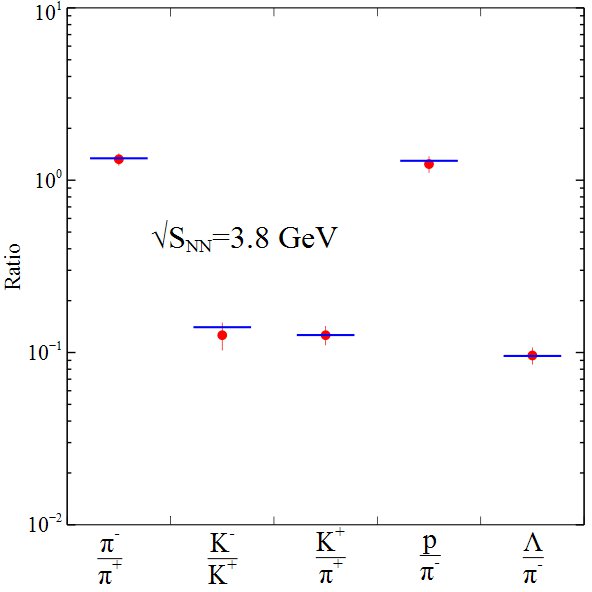}
}
\vspace*{2.2mm}
\centerline{
\includegraphics[width=84.mm]{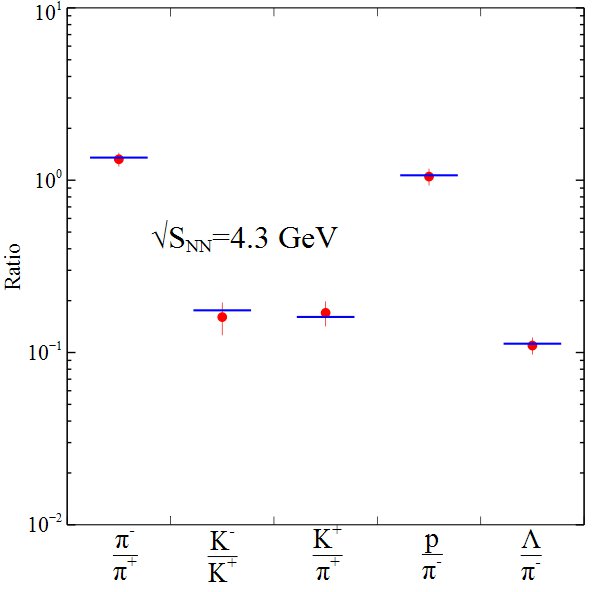}
}
 \caption{
 The particle-yield ratios described  by  the present multi-component HRGM. The best fit for   
 $\sqrt{s_{NN}} = 3.8$ GeV (AGS) is found for $T \simeq 84.0$ MeV, $\mu_B \simeq 
 618.5$ MeV,  $\mu_{I3} \simeq -13.3$ MeV,  yielding 
 $\chi^2/dof =0.56/2 =0.28$, whereas for 
 $\sqrt{s_{NN}} = 4.3 $ GeV (AGS) it is found for $T \simeq 95.3$ MeV, $\mu_B \simeq 
 586.7$ MeV,  $\mu_{I3} \simeq -16.2$ MeV, {with} $\chi^2/dof =0.35/2 =0.175$.  
}
  \label{fig1}
\end{figure}

\clearpage

\section{Novel Irregularities At Chemical FO}

In order to elucidate the irregularities at chemical FO we employ a multicomponent hadron 
resonance gas model (HRGM)  \cite{Oliinychenko:12,KAB:13}
which currently provides the best description of 
the observed hadronic multiplicities.  Hadron interactions are
taken into account via hard-core repulsion,
with different hard-core radii for pions, $R_{\pi}$, kaons, $R_K$, 
other mesons, $R_m$, and baryons, $R_b$. The best global fit of all hadronic multiplicities 
was found for $R_b$ = 0.2 fm, $R_m$ = 0.4 fm, $R_{\pi}$ = 0.1 fm, and $R_K$ = 0.38 fm   
\cite{KAB:13}. More details about the HRGM can be found in 
Refs.\ \cite{Oliinychenko:12,KAB:13,KAB:gs,NICAWP}. 

In the HRGM the full chemical potential of the $i$-th hadron 
species, $\mu_i \equiv Q_i^B \mu_B + Q_i^S \mu_S + Q_i^{I3} \mu_{I3}$, 
is expressed in terms of the corresponding charges $Q_i^K$  and their chemical potentials. 
For each collision energy the fit parameters are 
the temperature $T$, the baryonic chemical potential $\mu_B$, 
and the chemical potential of the third component 
of isospin $\mu_{I3}$, whereas the 
strange chemical potential $\mu_S$ is found from the condition of vanishing strangeness.
\begin{figure}[t]
\centerline{\includegraphics[width=84.mm]{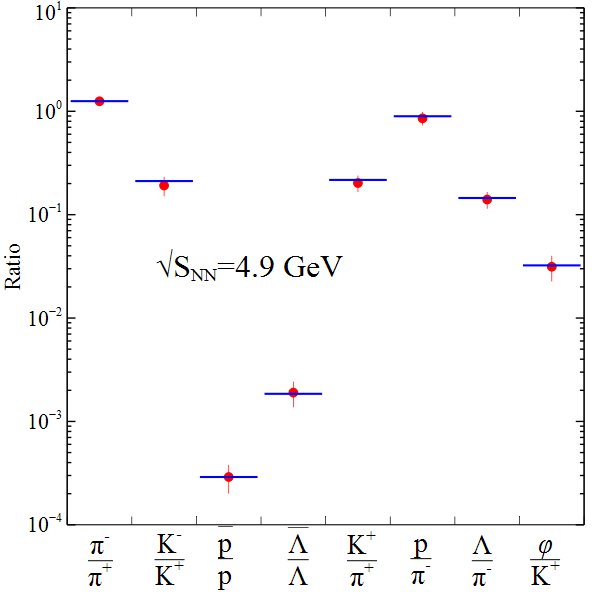}
}
\vspace*{2.2mm}
\centerline{
\includegraphics[width=84.mm]{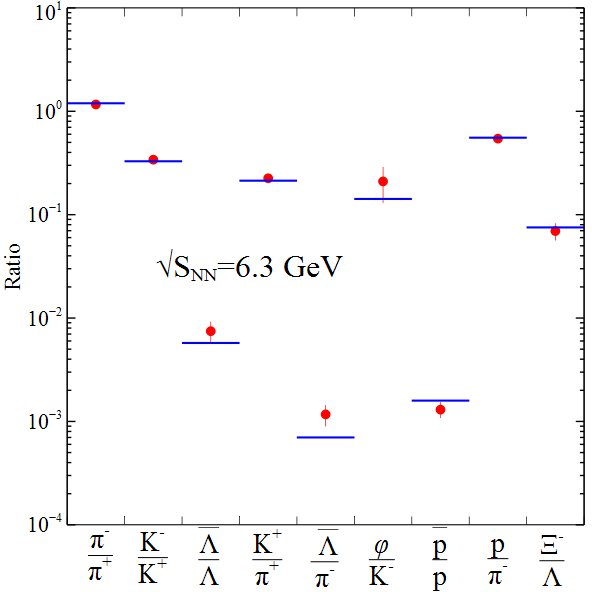}
}
 \caption{The same as in Fig.\ 1, but for $\sqrt{s_{NN}} = 4.9$ GeV (AGS). The fit gives  $T 
 \simeq 127.0$ MeV, $\mu_B \simeq 530.8$ MeV,  $\mu_{I3} \simeq -18.4$ MeV, and $\chi^2/
 dof =0.5/5 =0.11$, whereas  for  $\sqrt{s_{NN}} = 6.3 $ GeV (SPS) one finds   $T \simeq 
 130.1$ MeV, $\mu_B \simeq 430.4$ MeV,  $\mu_{I3} \simeq -15.6$ MeV, and $\chi^2/dof 
 =7.91/6 =1.32$.}
  \label{fig2}
\end{figure}

Another important feature of the employed HRGM is that the width of all hadron resonances is 
taken into account.  A thorough discussion of its role can be found in 
Ref.\ \cite{NICAWP}.
As usual, the effect of the resonance decay $Y \to X$ on the final 
hadron multiplicity is taken into account as $n^{fin}(X) = \sum_Y BR(Y \to X) n^{th}(Y)$, 
where $BR(X \to X)$ = 1 for the sake of convenience.  The masses, the widths, 
and the strong-decay branchings of all experimentally known hadrons were taken from 
the particle tables used by THERMUS \cite{THERMUS}. 

In this work  we use the data set of Ref.\ 
\cite{KAB:13}. At AGS energies ($\sqrt{s_{NN}}=2.7-4.9$ GeV or 
$E_{lab}=2.9-11.6$ GeV per nucleon) the available data have 
a  good energy resolution. 
However, for the beam energies 2, 4, 6, and 8 GeV per nucleon only a few data points are 
available.  They correspond  to the yields for pions \cite{AGS_pi1,AGSpi2}, for protons 
\cite{AGS_p1,AGS_p2}, and for kaons \cite{AGSpi2} (except for 2 GeV per nucleon).
$4\pi$--integrated data are also available for $\Lambda$ hyperons 
\cite{AGS_L} and for $\Xi^-$ hyperons (for 6 GeV per nucleon only) \cite{AGS_Kas}. 
However, as it  was argued in Ref.\ \cite{KABAndronic:05}, the data for $\Lambda$ and $\Xi^-$ 
should be recalculated for midrapidity. Therefore, instead of raw experimental data we used 
the corrected values from Ref.\ \cite{KABAndronic:05}. 
Furthermore, we analyzed  the data set at the highest 
AGS energy  \cite{AGS_p1,AGSpi2,AGS_L2} ($
\sqrt{s_{NN}}=4.9$ GeV or $E_{lab}=11.6$ GeV per nucleon). Similarly to 
Ref.\ \cite{KAB:13}, here we analyzed only the NA49 mid-rapidity data   
\cite{KABNA49:17a,KABNA49:17b,KABNA49:17Ha,KABNA49:17Hb,KABNA49:17Hc,KABNA49:17phi}, 
being the most difficult ones  to reproduce. 
Since RHIC high-energy data of different collaborations agree well with each other, we  
analyzed the STAR results for $\sqrt{s_{NN}}= 9.2$ GeV \cite{KABstar:9.2}, 
$\sqrt{s_{NN}}= 62.4$ GeV \cite{KABstar:62a}, $\sqrt{s_{NN}}= 130$ GeV 
\cite{KABstar:130a,KABstar:130b,KABstar:130c,KABstar:200a}, 
and 200 GeV \cite{KABstar:200a,KABstar:200b,KABstar:200c}.  

These experimental data were chosen because of their completeness  and 
consistency.  The fit criterion  is a minimization  of $\chi^2=\sum_i\frac{(r_i^{theor}-r_i^{exp})^2}
{\sigma^2_i}$, where $r_i^{theor}$ and $r_i^{exp}$ are, respectively,  theoretical and 
experimental  values of particle-yield ratios, $\sigma_i$ stands for the corresponding 
experimental error, and a  summation is performed over all independent experimental ratios. 
For an extended discussion of the independent hadron-multiplicity ratios see 
Refs.\ \cite{KAB:13,KABAndronic:05}.

Within the present HRGM, a global value of $\chi^2/dof \simeq 1.16$  
\cite{KAB:13}  is obtained  for 
111 independent multiplicity ratios measured at  AGS, SPS, and RHIC, at 
the 14 values of collision energy specified above.
Some results are shown in Figs.\ \ref{fig1} and \ref{fig2}. 
As one can see from {these figures}, AGS and SPS data are described with 
very high quality and, hence, using the fit parameters found
we were able to extract all thermodynamic quantities at chemical FO.

\begin{figure}[t]
\centerline{\includegraphics[width=88.mm]{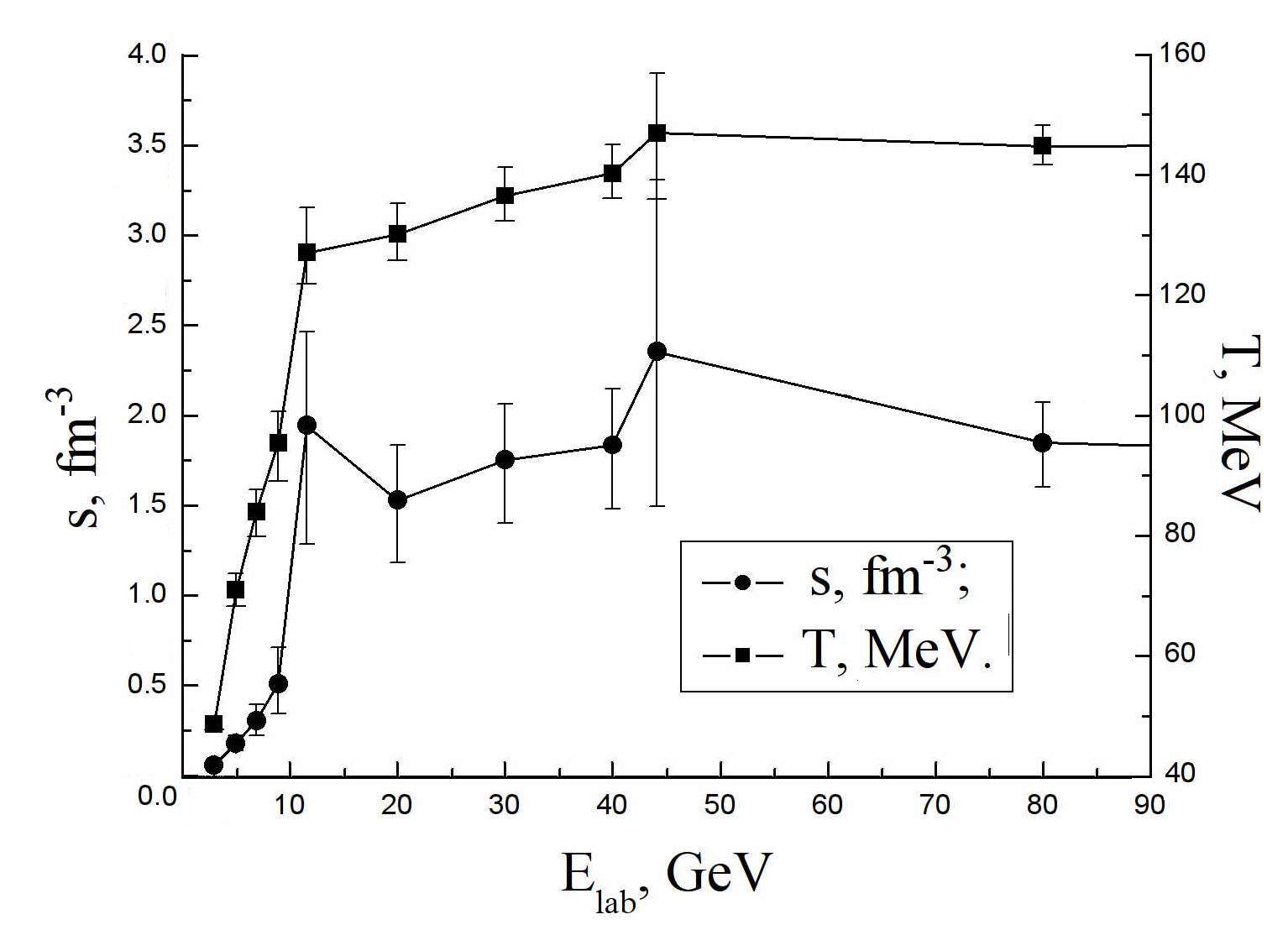}}
 \caption{Energy dependence of entropy density (circles) and temperature (squares) at 
 chemical freeze-out extracted from hadron multiplicities measured in 
 heavy-ion collisions at AGS ($E_{lab} < 15$ GeV),  SPS  ($E_{lab} > 15$ GeV), and RHIC 
 ($E_{lab} = 44$ GeV). 
}
  \label{fig3}
\end{figure}

\begin{figure}[t]
\centerline{~~\includegraphics[width=88.0mm,height=110mm]{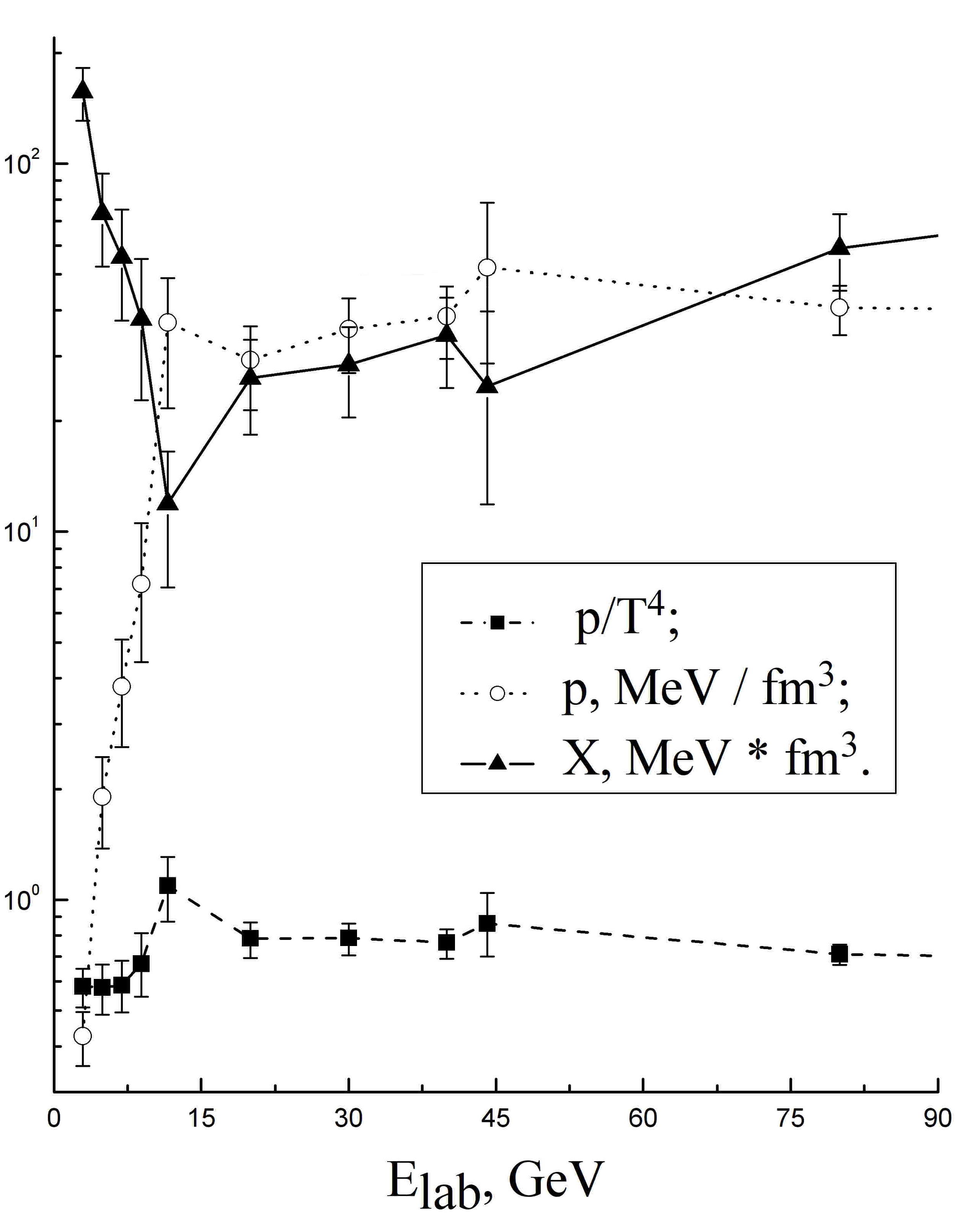}
}
 \caption{Energy dependence of the chemical freeze-out pressure (circles),  
 the effective number of degrees of freedom (squares),  
 and the generalized specific volume $X$ found by the present HRGM for the data measured 
 at AGS ($E_{lab} < 15$ GeV),  SPS  ($E_{lab} > 15$ GeV), and RHIC ($E_{lab} = 44$ GeV).
}
  \label{fig4}
\end{figure}

\begin{figure}[t]
\centerline{\includegraphics[width=88.mm,height=84mm]{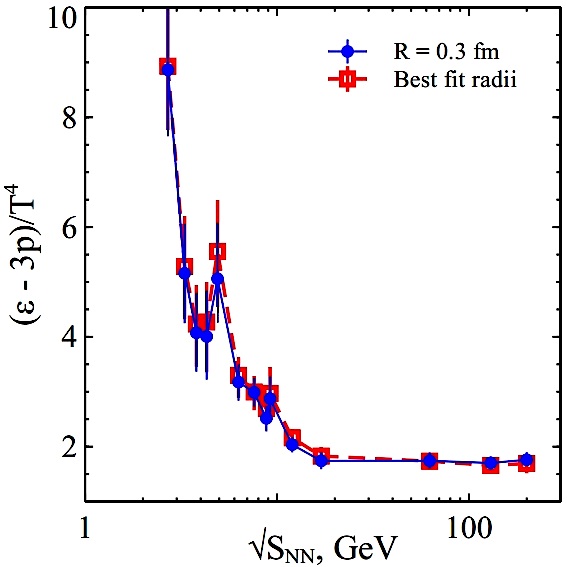}
}
 \caption{Energy dependence of the dimensionless trace anomaly $\delta$ at chemical FO   
 found within the HRGM with the same hard-core radius 0.3 fm for all hadrons 
 \cite{Oliinychenko:12}(circles) and with the set of hadronic hard-core radii found in  
 Ref.\ \cite{KAB:13} and employed here (squares). 
 The peak at $\sqrt{s_{NN}} = 4.9$ GeV (AGS) looks very similar to the 
 trace-anomaly peak reported by lattice QCD.  Lines are shown to guide the eye. 
}
  \label{fig5}
\end{figure}

The most remarkable results are depicted in Figs.\  \ref{fig3}, \ref{fig4}, and  \ref{fig5}. 
From Fig.\ \ref{fig3} one can see that the entropy density $s$ increases by a factor of four
in a range of laboratory energies per nucleon $E_{lab} \simeq 8.6 - 11.6$  GeV. At the same
time, the chemical FO temperature changes from 95 to 127 MeV and the
baryonic chemical potential $\mu_B$ drops only from 586 to 531 MeV. 
In other words, for a 30\% increase in the laboratory energy the ratio $s/T^3$ increases
by 70\%. A similar change can be seen in 
the effective number of degrees of freedom in the same energy range, cf.\ 
Fig.\ \ref{fig4}  for the chemical FO  pressure in units of $T^4$. Note that a similar 
(and somewhat stronger) rapid change in the number of effective degrees of freedom 
is observed in the most recent (former) version of the hadron resonance gas model
\cite{KAB:gs,NICAWP} (\cite{Oliinychenko:12,KABAndronic:05}).

As one can see from Fig.\ \ref{fig4}, at chemical FO a dramatic change is  
experienced by the so-called generalized specific volume 
$X=\frac{\varepsilon +p}{\rho_B^2}$, where
$\varepsilon$ is the energy density, $p$ the pressure,
and $\rho_B$ the baryonic charge density.  
It is remarkable that in all known examples of equations of state (EOS)
describing the QGP-hadron transition a  local minimum 
in the energy dependence of the $X$ values of matter described by the shock or generalized 
shock model
is observed right at the transition to the QGP, independently of whether this is a first-order 
phase transition \cite{KAB:89a,KAB:90,KAB:91,KAB:89} or a  
strong cross-over \cite{KAB:89a,KAB:89}.
Therefore, below we would like to reanalyze the generalized shock-adiabat model developed
in Refs.\ \cite{KAB:89a,KAB:90,KAB:91,KAB:89,KAB:89b} in order to interpret the above 
irregularities and to verify the other signals of mixed-phase formation.

Another irregularity, the behavior of the dimensionless trace anomaly $\delta = 
\frac{\varepsilon - 3p}{T^4}$, is shown in Fig.\ \ref{fig5}. {Here, one
observes a peak at $\sqrt{s_{NN}} = 4.9$ GeV. One could conservatively argue that
this peak is produced by only one data point and that its significance
should therefore be considered with caution. However, one can see that the error bars at  
$\sqrt{s_{NN}} = 4.9$ GeV and $\sqrt{s_{NN}} = 6.3$ GeV do not overlap.
Thus, we are inclined to argue that this peak is indeed physical. Further measurements
at various other collision energies in the immediate vicinity of 
$\sqrt{s_{NN}} = 4.9$ GeV could lend support to our argument.}

Note that the inflection point/maximum of the trace anomaly is traditionally used in lattice 
QCD to determine the pseudocritical temperature of the cross-over 
transition \cite{lQCD}.  Figure \ref{fig5} shows 
that a sharp peak of the trace anomaly exists at 
the center-of-mass energy $\sqrt{s_{NN}} = 4.9$ GeV and it is not a 
coincidence that it occurs right at this energy, since $\delta$ is directly related to the partial 
derivatives of the effective number of degrees of freedom $\frac{p}{T^4}$ as 
\begin{eqnarray}
\delta \equiv T \frac{\partial }{\partial T} \left( \frac{p}{T^4}\right) + \mu_B  \frac{\partial }{\partial 
\mu_B} \left( \frac{p}{T^4}\right) + \mu_{I3}  \frac{\partial }{\partial \mu_{I3}} \left( \frac{p}
{T^4}\right)\,.
\end{eqnarray}
Therefore, we face the intriguing possibility to relate the 
peak of the trace anomaly at chemical FO with its peak 
at the phase transition. In the next sections we will do this using 
the shock-adiabat model for central nuclear collisions. Also it is necessary to  
mention that, in order to see the peak of $\delta$ in Fig.\  \ref{fig5},   
it is rather  important to account for  the non-zero width of all hadronic resonances 
\cite{NICAWP}.

\section{Generalized Shock-Adiabat Model}

The generalized shock-adiabat model was developed in Refs.\
\cite{KAB:89a,KAB:90,KAB:91,KAB:89,KAB:89b} to extend the compression-shock model 
\cite{Mish:78,Horst:80,Kamp:83,Horst_PR:86,Barz:85} {to} 
regions of matter with anomalous
thermodynamic properties. Similarly to nonrelativistic hydrodynamics \cite{Landavshitz}, 
in the relativistic case matter is thermodynamically normal, if the quantity 
$\Sigma \equiv \left(\frac{\partial^2 p}{\partial X^2} \right)^{-1}_{s/\rho_B}$ 
is positive along the Poisson adiabat. Otherwise, for $\Sigma < 0$, the 
matter has thermodynamically anomalous properties. The sign of $\Sigma$ 
defines the type of allowed simple and shock waves: for $\Sigma > 0$  
rarefaction simple waves and compression shocks are stable. In the case of an 
anomalous medium, $\Sigma < 0$, 
compressional simple waves and rarefaction shocks  are stable. 
If both signs of $\Sigma$ are possible, then a more detailed investigation of the 
possible flow patterns is necessary \cite{KAB:89a,KAB:91}. In fact, all known pure 
phases have thermodynamically normal properties \cite{Landavshitz}, whereas 
anomalous properties may
appear at a first-order phase transition \cite{Zeldovich,Rozhd}, at its second-order
critical endpoint \cite{Landavshitz}, or for a fast cross-over \cite{KAB:89a}.

The compression-shock model of central nuclear collisions 
\cite{Mish:78,Horst:80,Kamp:83,Horst_PR:86,Barz:85} allows one to determine the 
initial conditions for the subsequent hydrodynamic evolution. Such a picture of the 
collision process, which neglects the nuclear transparency, can be reasonably well  
justified at intermediate collision energies per nucleon $1$ GeV $ \le E_{lab} \le $ 20 GeV.  
Its validity in this region of collision energies 
was once more demonstrated recently \cite{Satarov:11} within the
hydrodynamic model with a realistic  EOS. The 
direct comparison  of the results of the compression-shock model 
with those of the three-fluid model \cite{3fluid}
made in Ref.\ \cite{Satarov:11} shows that
at laboratory energies per nucleon up to 30 GeV this model can  
be used for quantitative estimates, while at higher energies it provides a 
qualitative description only.  Moreover, this comparison 
shows that at $ E_{lab} \le $ 30 GeV transparency 
effects are not so strong, at least in central nucleus-nucleus collisions.

\begin{figure}[t]
\centerline{\includegraphics[width=84.0mm,height=84mm]{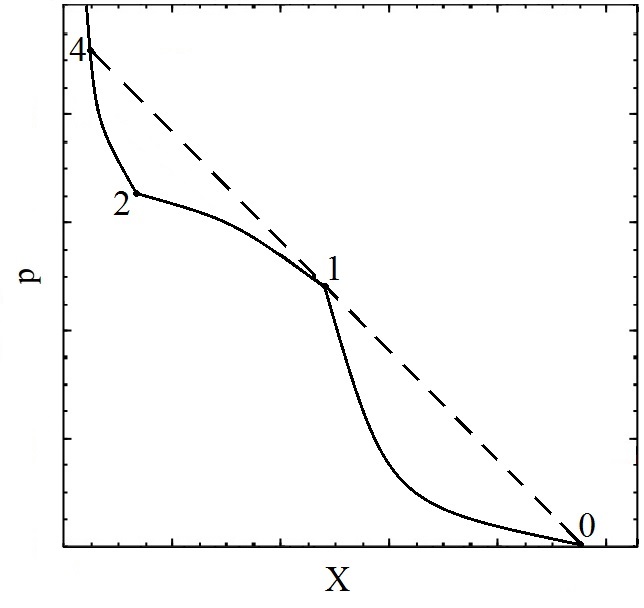}}
\vspace*{-0.11mm}
 \caption{Schematic picture of the compression RHT adiabat 0-1-2-4 (solid curve)  
 of W-kind \cite{KAB:89a} in the $X-p$ 
plane. It is calculated for an EOS with first-order phase transition 
as discussed in the text.  
The segments 0-1, 1-2, and 2-4 of the adiabat 
correspond to the hadronic, mixed, and QGP phases, respectively.   
Shock transitions into the region of states 1-2-4  are mechanically unstable.}
  \label{fig6}
\end{figure}

\begin{figure}[t]
\centerline{\includegraphics[width=84.0mm,height=84mm]{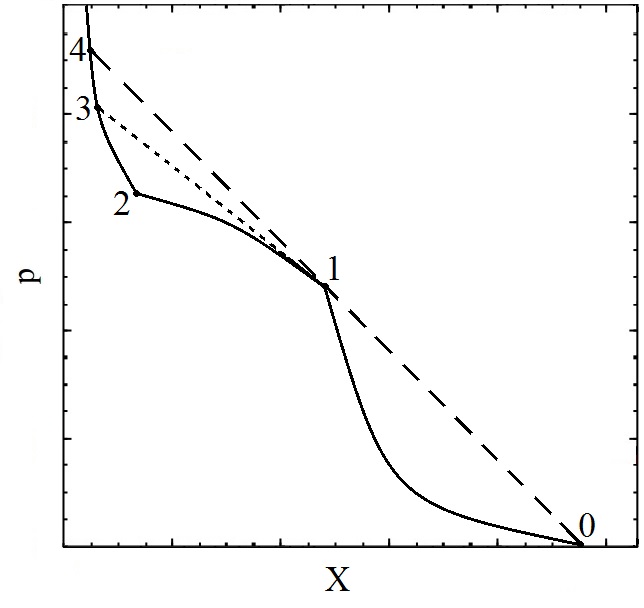}}
\vspace*{-0.11mm}
 \caption{Schematic picture of the generalized shock adiabat 0-1-2-3-4 (solid curve)  
 of W-kind \cite{KAB:89a} in the $X-p$ plane.  
The segments 0-1, 1-2, and 2-4 of the adiabat 
correspond to the hadronic, mixed, and QGP phases, respectively.   
Shock transitions into the region of states 1-2-3-4 are not allowed.
The generalized solutions corresponding to these segments are discussed in the text.
The dashed line 1-3 is the tangent line in the left vicinity of the point 1. }
  \label{fig7}
\end{figure}

In the center-of-mass frame of the two colliding nuclei the 
initial state of the collision can be considered as a hydrodynamic Riemann 
problem of an initial discontinuity. For normal media this kind of initial discontinuity leads to 
an appearance of two compression shocks that move in opposite directions 
toward the vacuum, leaving high-density matter at rest behind the shock fronts.
The thermodynamic  parameters $X, p, \rho_B$  of this compressed matter  
are related by the Rankine-Hugoniot-Taub (RHT) adiabat \cite{Landavshitz} 
with the parameters of uncompressed matter in the state $(X_0, p_0,\rho_{B0})$, 
\begin{eqnarray}\label{EqI}
\rho_{B}^2 X^2 -  \rho_{B0}^2 X^2_0 = (p - p_0)\left( X + X_0  \right) \,. 
\end{eqnarray}
This equation follows from the usual hydrodynamic conservation laws of energy, momentum, 
and baryonic charge across the shock front. 
The variable $X$ is convenient, since with its help the conserved baryonic current 
can be expressed as  
%
%
$j^2_B = - \frac{ p - p_0}{ X - X_0 }$,
%
%
i.e., in the $X-p$ plane the state existing behind the shock front is 
given by the intersection point 
of the RHT adiabat (\ref{EqI}) and the straight line with the slope $j^2_B$ 
known as the Raleigh line. In order to solve Eq.\ (\ref{EqI}) one needs to know the EOS.  

Traditionally,  it is believed that the compression-shock model requires immediate 
thermalization. However, we would like to stress that the 
hydrodynamic conservation laws should be obeyed irrespectively
of whether matter is in thermodynamic equilibrium or not. Moreover, the propagation of the 
compression shock (or a more complicated hydrodynamic solution) requires 
a finite formation time $t_{form}$ which in the center-of-mass frame of 
the collision is essentially larger than the propagation of light along the 
Lorentz-con\-tracted nucleus of diameter $d$, i.e., 
$$ {t_{form} > \frac{d}{\gamma_0}= \frac{d}{\sqrt{1 + E_{lab}/(2 m_N)}}}$$ 
(here, $m_N $ is the mean nucleon mass). 
More accurate estimates for the time of the collision can be found in 
Ref.\ \cite{KAB:89a}, but the order of magnitude of the above estimate 
should be correct. Hence, for $E_{lab}= 10$ GeV and for $E_{lab}= 30$ GeV 
one finds $t_{form} > d/2.5 \simeq 4.7$ fm and $t_{form} > d/4.1 \simeq 2.9$ fm,
respectively,
where we used $d = 12$ fm. In other words, this formation time is rather long on the scale of 
the strong interaction and, hence, after this time the major part of matter 
formed in the central zone of the collision has a good chance to become
equilibrated, even if it was initially out of equilibrium. Consequently, 
if only a small amount of energy dissipates during the formation time, 
it is reasonable to assume that the 
formed matter can be described by the usual hydrodynamic conservation laws for 
a perfect fluid \cite{Landavshitz}.

In the compression-shock model the laboratory energy per nucleon is
\begin{eqnarray}\label{EqII}
E_{lab} = 2 m_N \left[ \frac{(\varepsilon+p_0)(\varepsilon_0+p)}{(\varepsilon+p)(\varepsilon_0+p_0)} - 1 \right] \,.
\end{eqnarray}
A typical example for the shock adiabat is shown in  
Fig.\ \ref{fig6}. As one can see from this figure the shock adiabat in the pure 
hadronic and QGP phases exhibits the typical (concave) behavior for a normal medium, while 
the mixed phase (the region 1-2) in Fig.\ \ref{fig6} has a convex shape 
which is typical for matter with anomalous properties. 
Until  now there is no complete understanding why in a phase-transition or cross-over region 
matter exhibits anomalous thermodynamic properties. In pure gaseous or liquid phases the 
interaction between the constituents at short distances is repulsive and, hence, at high  
densities the adiabatic compressibility of matter 
$- \left( \frac{\partial X}{\partial p}\right)_{s/\rho_B}$ usually decreases for 
increasing pressure, i.e.,  
$ \left(\frac{\partial^2 p}{\partial X^2} \right)^{-1}_{s/\rho_B} =  \Sigma >0$. In the mixed phase 
there appears another possibility to compress matter: by converting the less dense phase 
into the more dense one.  As it was found for several EOS with a first-order phase transition 
between hadronic gas and QGP, the phase transformation leads to an increase of the
compressibility in the mixed phase at higher pressures, i.e., to 
anomalous thermodynamic properties. 
The hadronic phase of the aforementioned EOS was described by the Walecka model 
\cite{Walecka} and by a few of its more realistic phenomenological generalizations 
\cite{KAB:91,Satarov:11,Zimanyi}.
The appearance of anomalous thermodynamic properties for a fast cross-over can 
be understood similarly, if one formally considers the cross-over states as a kind of 
mixed phase (but without sharp phase boundary), in which, however,  
none of the pure phases is able to completely dominate.

\begin{figure}[t]
\centerline{\includegraphics[width=84.0mm,height=55mm]{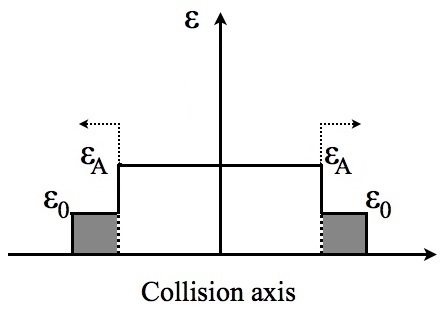}}
\centerline{\includegraphics[width=84.0mm]{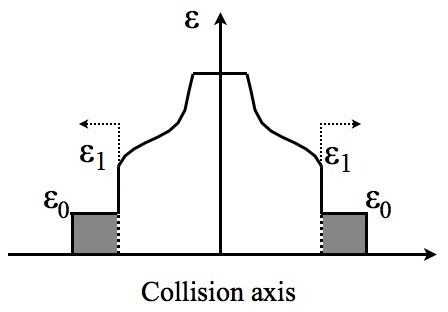}}
 \caption{Upper panel: Sketch of the energy density profile for a collision of two nuclei (grey 
 areas) that corresponds to stable shock transitions (segments 0-1 and above point 4) of the
generalized shock adiabat of Fig.\ \ref{fig7}. The dashed arrows show the 
direction of shock propagation.
Lower panel: Same as in the upper panel, but for the segment 1-2 of the
generalized shock adiabat of Fig.\ \ref{fig7}. 
Two shocks between the states $\varepsilon_0 \rightarrow \varepsilon_{1}$ are 
followed by compressional simple waves. 
}
  \label{fig8}
\end{figure}

\begin{figure}[t]
\centerline{\includegraphics[width=84.0mm,height=55mm]{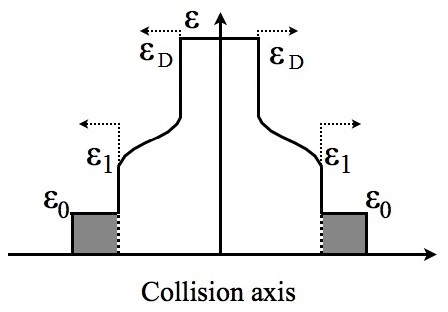}}
\centerline{\includegraphics[width=84.0mm]{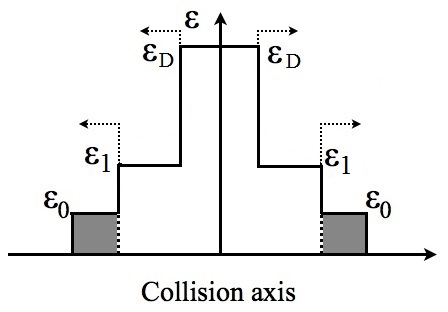}}
 \caption{Upper panel: Sketch of the energy density profile for a collision of two nuclei (grey areas) that corresponds to the states belonging to the segment 2-3 of the
generalized shock adiabat of Fig.\ \ref{fig7}. The dashed arrows show the 
 direction of shock propagation. Between each pair of shocks exists a simple wave. 
Lower panel: Same as in the upper panel, but for the segment 3-4 of the
generalized shock adiabat of Fig.\ \ref{fig7}. In this case on each side there is a double shock 
 between the states $\varepsilon_0 \rightarrow \varepsilon_{1}$ and  
 $\varepsilon_{1} \rightarrow \varepsilon_{D}$.
}
  \label{fig9}
\end{figure}


\begin{figure}[t]
\centerline{\includegraphics[width=84.0mm,height=63mm]{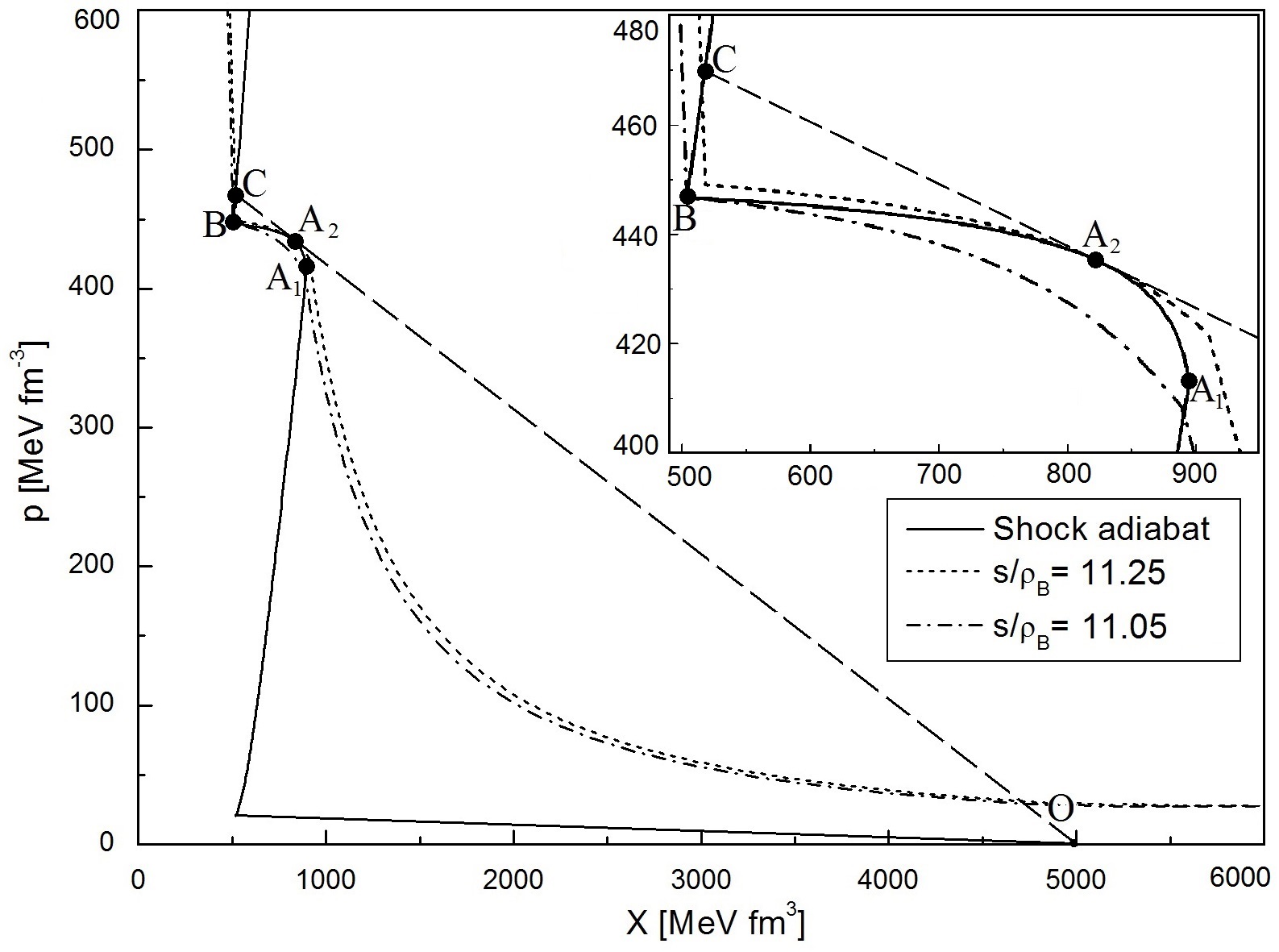}}
\vspace*{0.11mm}
 \caption{The compression RHT adiabat OA$_2$BC (solid curve) of W-kind in the $X-p$ 
plane. It is calculated for the EOS with first-order phase transition discussed in the text.  
The segments OA$_1$, A$_1$B, and BC of the adiabat 
correspond to the hadronic, mixed, and QGP phases, respectively.   
Shock transitions into the region of states A$_2$BC are mechanically unstable. The tangent 
point A$_2$ to the shock adiabat is the Chapman-Jouguet point \cite{Landavshitz}.
The dotted and dashed-dotted curves show the Poisson adiabats with 
values of entropy per baryon specified in the legend. See insert figure for details.}
 \label{fig10}
\end{figure}

From Fig.\ \ref{fig6} one sees that the presence of anomalous matter  
leads to mechanically unstable parts of the RHT adiabat (segment 1-2-3 in Fig.\ \ref{fig6}), 
which include states in the mixed and the QGP phases. 
This is a model of W-kind \cite{KAB:91,Satarov:11,Walecka} and   
its RHT adiabat in the instability  
region should be replaced by the generalized shock adiabat (see Fig.\ \ref{fig7}) 
\cite{KAB:89a,KAB:90,KAB:91}. 
In this case, in the region of instability 1-2-3 the shock wave has to be replaced 
by the following hydrodynamic solution \cite{KAB:89a} of the generalized shock adiabat 
shown in Fig.\ \ref{fig7}: a shock between 
states 0 and 1 (see the upper panel of  Fig.\ \ref{fig8}), followed by a compressional 
simple wave (see the lower panel of Fig.\ \ref{fig8}) for the segment 1-2
of the generalized 
shock adiabat of Fig.\ \ref{fig7}; at higher energies (segment 2-3 of the generalized shock 
adiabat of Fig.\ \ref{fig7}) this solution converts into two 
compressional shocks and a compressional simple wave moving between them shown in the 
upper panel of Fig.\  \ref{fig9}; 
at even higher energies (segment 3-4 of the generalized shock adiabat of Fig.\ \ref{fig7})
there exist two compressional shocks following one after the other \cite{KAB:89a,Mish:78}, 
i.e., a double shock wave as shown in the lower panel of Fig.\ \ref{fig9}.  
Note that the double-shock solution was already suggested a 
long time ago \cite{Bethe:42}.

A similar situation occurs in the case of a fast cross-over or for the RHT adiabat of the W-kind 
shown in Fig.\  \ref{fig10}
(see also Figs.\ 3 and 4 in Ref.\ \cite{KAB:89a} for more details) which has the 
Chapman-Jouguet point A$_2$ (see Fig.\ \ref{fig10}).
From Fig.\ \ref{fig10} one sees that the presence of anomalous matter  
leads to mechanically unstable parts of the RHT adiabat 
(segment $A_2BC$ in Fig.\ \ref{fig10}), which include states in the mixed and the 
QGP phases. 
The only difference with the case of the unstable RHT adiabat of Fig.\ \ref{fig6} is that
the double shock does not appear due to an instability, while 
the other solutions shown in Fig.\ \ref{fig8} and in the upper panel of 
Fig.\ \ref{fig9} remain the same.

It is necessary to stress that all these generalized
hydrodynamic solutions are proven to be mechanically stable \cite{KAB:89a,KAB:89,KAB:89b} 
and they were constructed in such a way that all neighboring shocks and simple waves are 
moving with the same velocity in the rest frame of  their common fluid. In case of the double 
shock the inner shock 
between the initial state 1 and the final state belonging to the segment 3-4 of the 
generalized shock adiabat of Fig.\ \ref{fig7}
propagates in the medium 1 slower than the shock between the states 0 and 1 
\cite{KAB:89a,KAB:89,KAB:89b}. 

Shock transitions to mechanically unstable regions are accompanied by a
thermodynamic instability, i.e., the entropy in such transitions decreases, while 
the collision energy grows \cite{Zeldovich,Rozhd,KAB:89a}. 
At the same time the mechanical stability condition of the generalized shock adiabat 
always leads to 
thermodynamic stability of its flows. Or in other words, along the correctly constructed  
generalized shock adiabat the entropy cannot decrease  \cite{KAB:89b}. 
Among the possible solutions mentioned above an important role is played by 
the combination of a
shock wave between the states O and A$_2$, followed by a simple wave starting in  
the state A$_2$ and continuing to states located at the boundary between the mixed phase  
and the QGP \cite{KAB:89a}. The energy-density profile of this solution is shown in the lower 
panel of Fig.\ \ref{fig8}. For such a solution the entropy does not grow
with collision energy, because the whole entropy is generated by a shock OA$_2$ to
the Chapman-Jouguet point A$_2$ (see Fig.\ \ref{fig10}).
This  means that by increasing the collision energy one generates more
highly compressed states which, however, have the same value of $s/\rho _B$.

Based on this solution a signal for mixed-phase formation was suggested, provided  
that this instability of a W-kind model exists \cite{KAB:89a,KAB:90,KAB:91}. 
The important physical consequence of such an instability
is a plateau in the collision-energy dependence of the total number of pions per baryon
produced in a 
nuclear collision, i.e., $\rho_\pi^{tot}/\rho_B (E_{lab}) \simeq const$  \cite{KAB:90,KAB:91}, 
provided that entropy is conserved during the subsequent hydrodynamic expansion. 
Since the total pion multiplicity consists of thermal pions and the ones which appear 
from decays of hadronic resonances, in case of the RHT adiabat the  
number of thermal pions per baryon $\rho_\pi^{th}/\rho_B$ should also 
exhibit a plateau or a plateau-like behavior with a small negative slope  
as a function of collision energy. 

\begin{figure}[t]
\centerline{\hspace*{7.7mm}\includegraphics[width=96.0mm,height=77mm]{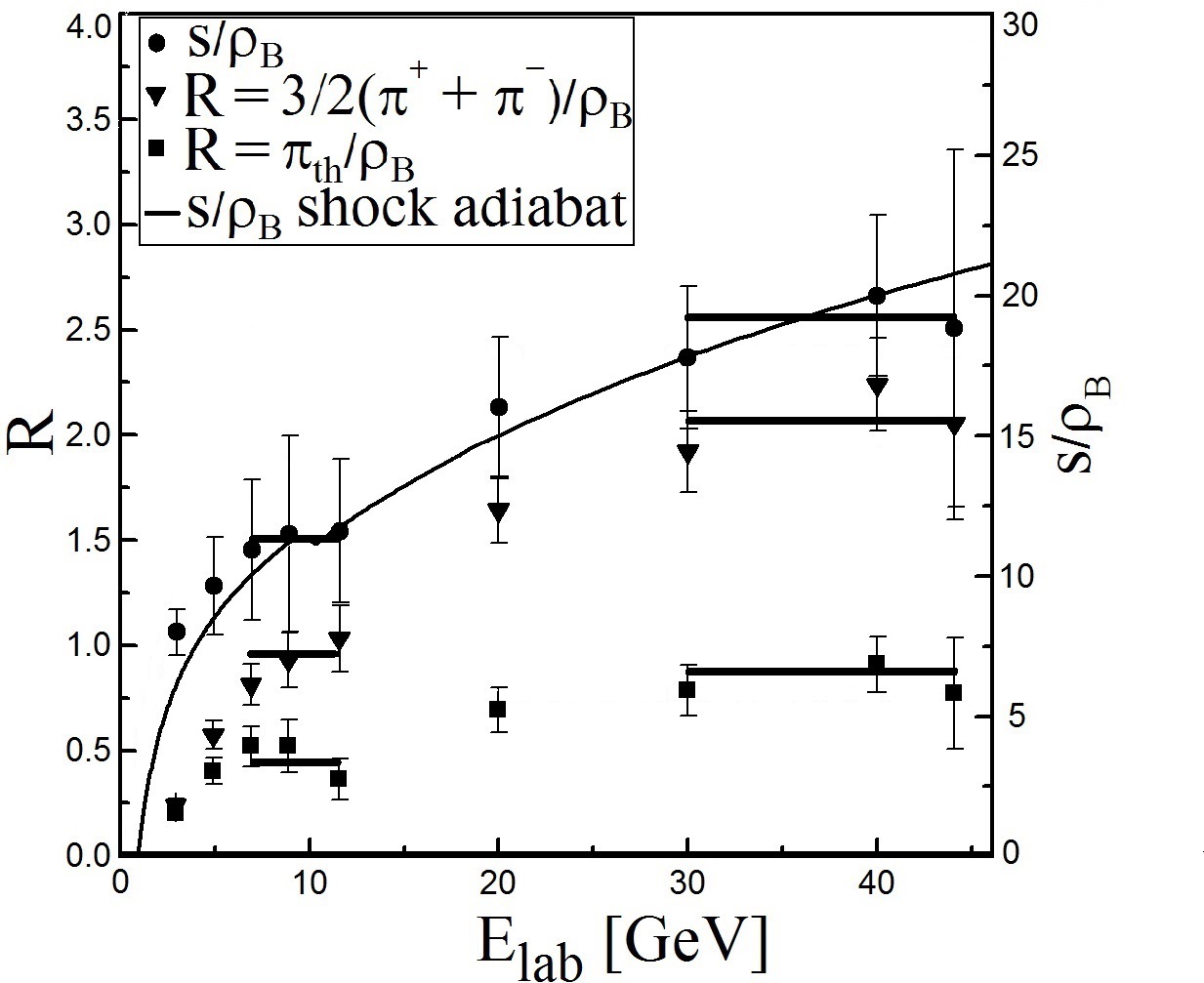}
}
 \caption{Energy dependence of the entropy per baryon (circles), of the thermal pion 
 multiplicity per baryon (squares), and of the total pion multiplicity per baryon (triangles) 
 found at chemical freeze-out within the present HRGM. The horizontal bars are found by 
 minimizing $\chi^2/dof$ (see text). 
 The solid curve corresponds to the RHT adiabat shown in Fig.\  \ref{fig10}.
}
  \label{fig11}
\end{figure}

In Fig.\ \ref{fig11} one can see {indications
for plateaus in the entropy per baryon 
and in the total and thermal pion number per baryon as functions of collision energy}
at the laboratory energies $E_{lab} \simeq 6.9 - 11.6$ GeV, i.e., 
{in the range}
where the other irregularities depicted in Figs.\ \ref{fig3}-\ref{fig5} occur. 
{We admit that the error bars for $s/\rho_B$ and the total number of pions
are still too large to draw definite conclusions. 
However, the thermal pion number per baryon clearly exhibits a
plateau and, as discussed above, even a small decrease at $E_{lab} \simeq 11.6$ GeV. We would like to stress that}
all quantities shown in Figs.\ \ref{fig3}-\ref{fig5} and \ref{fig11}
were found at chemical FO within the most realistic HRGM  
with multicomponent hard-core repulsion 
\cite{KAB:13}, which not only successfully describes 111 independent 
hadron-multiplicity ratios measured for  
center-of-mass energies  $\sqrt{s_{NN}} = $ 2.7 -- 200 GeV, but also correctly  
reproduces the energy dependence of the Strangeness Horn 
with a $\chi^2/dof \simeq 7.5/14$. 
 
It is, of course, possible that the anomalous properties of the mixed phase do not generate 
mechanical instabilities for shock transitions to this phase \cite{KAB:89a}, 
like it was found for the Z-kind  
hadronic EOS \cite{Zimanyi}. Nevertheless, it was argued that even in the latter case 
a plateau-like structure in the entropy per baryon, and, hence, in the thermal pion 
multiplicity per baryon, should be seen \cite{KAB:89a,KAB:90,KAB:91}.  

Now we are at a position 
to determine the parameters (individual heights for the same width) of the plateaus in the 
ratios $s/\rho_B$, $\rho_\pi^{th}/\rho_B$, and $\rho_\pi^{tot}/\rho_B$ shown in Fig.\ \ref{fig11}.
We investigated a few different schemes, but came to the conclusion that a
{three}-parameter fit is the most reliable and simple one. 
Since we are searching for a plateau it is 
clear that its height $R_A$ should be the same for a given quantity 
$A \in \{s/\rho_B; \rho_\pi^{th}/\rho_B; \rho_\pi^{tot}/\rho_B\}$. The width of all plateaus 
in the collision energy should also be the same, since they are generated by the 
same physical mechanism. Let $i_0$ denote the beginning of plateau, 
while  $M$ denotes its width. Then one has to minimize 
\begin{equation}
\chi^2/dof= \frac{1}{3M-3}\sum_A \sum_{i=i_0}^{i_0+M-1}
\left(\frac{R_A-A_i}{\delta  A_i} \right)^2 
\label{Plat}
\end{equation}
for all possible values of $i_0$ and $M > 1$. Here the subscript $i$ counts the data 
points $A_i$ to be described, whereas $\delta A_i$ denotes the error of the
corresponding quantity $A_i$. We assume that the plateaus are correlated to each other, 
if $\chi^2/dof$ is essentially smaller than 1. 
Also from the practical point of view it is necessary to find the  set of  
maximally correlated plateaus for future experiments.
The height of each plateau $R_A$ is found by minimizing $\chi^2/dof$ in
Eq.\ (\ref{Plat}) with respect to $R_A$ and one obtains
\begin{eqnarray}
\label{V}
R_A &= &\sum_{i=i_0}^{i_0+M-1}\frac{A_i}{\left(\delta A_i \right)^2}\biggl/
\sum_{i=i_0}^{i_0+M-1}\frac{1}{\left(\delta A_i \right)^2} \,.
\end{eqnarray}
As one can see from the table below minimal values of $\chi^2/dof < 0.2$ are 
reached for $M=2$, but these are not the widest plateaus. There exist two sets 
for $M=3$ with $\chi^2/dof \simeq 0.53$ for the low-energy plateaus 
(at $E_{lab} \simeq 6.9-11.6$ GeV)
and with $\chi^2/dof \simeq 0.34$ for the high-energy plateaus 
(at $E_{lab} \simeq 30-40$ GeV). Precisely these sets are 
depicted in Fig.\ \ref{fig11}, since we believe 
that the high-energy set at $E_{lab} \simeq 20-40$ GeV  
with the width $M=4$ should not be taken into account because  its 
value of  $\chi^2/dof \simeq 0.87$
is too close to 1 and, therefore, such plateaus are not strongly correlated 
even for these huge error bars. 

\begin{table}[h]
\noindent\caption{Results of the 3-parameter fit.}\vskip-3mm\tabcolsep3.2pt
\begin{center}
\begin{tabular}{|c| c| c| c| c| c|}
\hline  
&                                                    \multicolumn{5}{c|}{Low-energy minimum}  \\ \hline             
  ~$M$~  & ~$i_0$~ &  ~$R_{s/\rho_B}$~ &  ~$R_{\rho_\pi^{th}/\rho_B}$~ & ~$R_{\rho_\pi^{tot}/\rho_B}$~ & ~$\chi^2/dof$~  \\\hline 
     2     &    3    &      11.13        &        0.520                  &                 0.857                  &     0.178 \\\hline 
      3     &    3    &      11.31        &       0.461                   &                 0.892                  &     0.531  \\\hline  
 4     &    2    &      10.56        &       0.433                   &                 0.725                  &     1.649   \\\hline 
   5     &    2    &      11.54        &       0.470                   &                 0.848                  &     4.455   \\\hline    
 &                                           \multicolumn{5}{c|}{High-energy minimum}                                                    \\ \hline
   
     2     &     8     &      19.81       &         0.882                 &                2.204                    &      0.128    \\\hline
  3     &   7     &      18.78       &         0.835                 &                2.058                    &      0.340           \\\hline 
  4     &   6     &      17.82      &         0.779                 &                1.878                    &      0.871   \\\hline 
    5  & 5    &     16.26       &         0.648                 &                1.621                    &      3.721      \\\hline          
\end{tabular}
\end{center}
\end{table}

\vspace*{0.11mm}

Although in Fig.\ \ref{fig11} a clear plateau is seen in the thermal pion number per baryon 
while other quantities exhibit quasiplateaus, they are important since their strong 
correlation with the plateau  in  the thermal number of pions per baryon 
 found here allowed us to determine their common width in 
the laboratory energy.

\section{New signals of QGP formation}

Here we suggest a couple of new signals indicating a boundary between 
the mixed phase and the QGP. 
First, a local minimum of the generalized specific volume $X$ 
at chemical FO as a function of the collision energy (see Fig.\ \ref{fig4}),
signifies the existence of mechanical instabilities inside the mixed phase. 
Note that all stable RHT adiabats of Z-kind and all unstable RHT adiabats of W-kind and 
the corresponding  generalized shock adiabats with a QGP EOS 
of MIT-Bag model type \cite{MITBag} studied in Refs.\
\cite{KAB:89a,KAB:90,KAB:91,Satarov:11} demonstrate exactly the same behavior. 
The physical origin for such a behavior is that for an increase in collision energy  
the entropy per baryon  
and the temperature of the formed QGP (being a normal medium) increase as well, while the 
baryonic density and the baryonic chemical potential are  steadily decreasing. 
Hence in the QGP phase the variable $X \equiv (T s/\rho_B + \mu_B)/\rho_B$ grows, if the 
collision energy increases. Intuitively, such a dependence seems to be true for other QGP 
EOS, if they correspond to a normal medium. On the other hand, the behavior of the variable 
$X$ inside the mixed phase with anomalous properties is opposite 
and it does not depend on the stability or instability of the shock transitions 
to this region \cite{KAB:89a,KAB:90,KAB:91}. On the basis of these arguments one  
can understand the reason why the boundary of the mixed phase and QGP corresponds to a  
local minimum of the $X$ variable along the RHT adiabat (shown in Fig.\ \ref{fig10}) 
or generalized shock adiabat  and why it is also a minimum of $X$ as function of 
collision energy \cite{KAB:89a,KAB:90,KAB:91,Satarov:11}.

In case of unstable shock transitions to the mixed phase, the unstable part of the mixed
phase (segment A$_2$B in Fig.\ \ref{fig10}) should be replaced by the Poisson adiabat 
passing through the point A$_2$ (the dotted curve shown in Fig.\ \ref{fig10}).  
Consequently, if the matter formed in a collision expands isentropically after the 
shock OA$_2$ disappears, then it can be shown that for 
the chemical FO pattern depicted in Figs.\ \ref{fig3} and \ref{fig4} the minimum of the variable 
$X$ of the initial state
corresponds to a minimum of this variable at chemical FO, i.e., $\min\{X (E_{lab})\}$ 
corresponds to $\min\{X^{FO} (E_{lab})\}$. Indeed, the final states of the 
isentropic expansion belong to the Poisson adiabat at which 
$s/\rho_B = s^{FO} V^{FO}/ (2A) = const$. Here the entropy density $s^{FO}$ and the system 
volume $V^{FO}$ are taken at chemical FO, while the total number of baryons in
an A+A collision is $2A$. At chemical FO temperatures $T$ below 150 MeV, the hadronic  
EOS can be safely represented as a mixture of ideal gases of massive  pions and nucleons, 
i.e., its pressure $p \simeq T (\rho_B + \rho_\pi) $ and energy density $\varepsilon \simeq 
(m_N + 3/2\,T) \rho_B + (m_\pi+3/2\,T)\rho_\pi $ can be represented via the 
density of nucleons $\rho_B$ and density of pions $\rho_\pi$ (here $m_N$ ($m_\pi$) 
is the nucleon (pion) mass).  With the help of this EOS the variable $X$ at chemical FO 
can be cast as $X^{FO} \simeq [m_N+m_\pi + 5/2 \, T(1+ \rho_\pi/\rho_B)] V^{FO}/ (2A)$.  
From Fig.\ \ref{fig11} one sees that constant values of $s/\rho_B$ 
in the range of $E_{lab} =  6.9-11.6$ GeV correspond to a nearly 
constant ratio $\rho_\pi/\rho_B \simeq 0.5$ and, hence, one can write  
$X^{FO} \simeq [m_N+m_\pi + 3.75 \, T] V^{FO}/ (2A) $ for these energies.  
Since for these energies the entropy density changes from 0.3 fm$^{-3}$ to 1.944 fm$^{-3}$, 
while the chemical FO temperature changes from 84 MeV to 127 MeV, it is clear that 
approximately one can write $X^{FO} \simeq [m_N+m_\pi + 358 \,{\rm MeV}] V^{FO}/ (2A)$ 
and, hence, the value $X^{FO} s^{FO} \simeq [1536 \, {\rm  MeV} ] V^{FO} s^{FO}/ (2A) = 
const$. 
A direct numerical check shows that for the chemical FO data belonging to the laboratory
energy range $E_{lab} =  6.9-11.6$ GeV one obtains 
$X^{FO} s^{FO} \simeq 16.9; 19.3; 23.1$, which means that such a relation is 
valid with relative deviations  $-12$\% and $+21$\%.
Note that the relation $X^{FO} s^{FO}/ [m_N+m_\pi + 3.75 \, T] \simeq const.$ gives us the 
values $15; 16.5; 17.9$, i.e., it is fulfilled with relative errors $-9.1$\% and $+8.5$\% 
in this energy range and these estimates validate the use of our EOS.  
Using these arguments, we conclude that with reasonable accuracy one can establish the 
relation $X^{FO} \sim  V^{FO} \sim 1/ s^{FO}$ for the final states which belong to the Poisson 
adiabat and, therefore, the growth of entropy density (see Fig.\ \ref{fig3}) and the decrease of
the variable $X$ shown in Fig.\ \ref{fig4} are directly related to each other. 

The same treatment can be applied to higher energies. In this case one has to write $X^{FO} 
\simeq [m_N+m_\pi + 5/2 \, T^{FO}(1+ \rho_\pi^{FO}/\rho_B^{FO})] s^{FO}/\rho_B^{FO}\, / 
s^{FO}$ and account for the 
fact that the ratios $s^{FO}/\rho_B^{FO}$ and $\rho_\pi^{FO}/\rho_B^{FO}$ are increasing with 
collision  
energy, while the chemical FO temperature and entropy density are almost constant for  
$E_{lab} > 11.6$ GeV (see Fig.\ \ref{fig3}). Hence, in this case one can write $X^{FO} \sim
s^{FO}/\rho_B^{FO}\sim s^{init}/\rho_B^{init}$, 
where we used the entropy conservation $s^{init}/\rho_B^{init} = s^{FO}/\rho_B^{FO}$ during 
the expansion of the matter formed after the disappearance of the shock wave (initial matter).
Therefore, for laboratory energies above 11.6 GeV the variable $X^{FO}$ should 
increase with the collision energy, and this is a reflection of the growth of the initial values of 
the $X$ variable, when the generalized shock adiabat reaches
states inside the QGP.

Accounting for the above estimates, we conclude that the local minimum of $X^{FO}$ 
is related to the minimum of the variable $X$ 
on the generalized shock adiabat existing at the boundary between the mixed phase and 
the QGP. Moreover, the above estimates show that the minimum of the function
$X^{FO} (E_{lab})$ 
corresponds to the minimum of the chemical freeze-out volume $V^{FO} (E_{lab})$, 
reported in Ref.\ \cite{KABAndronic:05} and reanalyzed recently in Ref.\ 
\cite{Oliinychenko:12}. Thus, we find that the minimum of $V^{FO} (E_{lab})$ is 
generated by the unstable part of the RHT adiabat which is close to the boundary of 
the mixed with the QGP phase, i.e., it is another signal of QGP formation.

In the above treatment we did not consider the kinetic FO process of  
hydrodynamic flows and did not account for the possible entropy growth in 
a kinetic FO shock \cite{KSH:96}.
The reason is that the kinetic FO shock generates not more than 
2\% of the entropy \cite{KSH:99} and, hence, we can safely neglect such an increase 
during kinetic FO and can further assume that entropy conservation holds.

Note that these conclusions were also verified numerically for the shock adiabat shown in 
Figs.\ \ref{fig10} and \ref{fig11} and, hence, it is appropriate to present here the 
employed EOS. The  hadron gas pressure used in the present work accounts for 
the mesonic and the (anti)baryonic states which are described by the masses $m_M$, $m_B$ 
and by the temperature-dependent numbers of degrees of freedom \cite{KABPoS12}
\begin{eqnarray}
\label{EqIV}
p_H &=& \left[  C_B T^{A_B}e^{-\frac{m_B}{T}} \cdot2\cosh\left({\mu}/{T}\right) 
\right. \nonumber \\
      &+ &    \left.   
        C_M T^{A_M}e^{-\frac{m_M}{T}}\right]\cdot e^{-\frac{p_HV_H}{T}}\,.        
\end{eqnarray}
This EOS  accounts for the short-range repulsion introduced via the excluded volume $V_H=
\frac{4}{3}\pi R_H^3$  (with $R_H = 0.3$ fm) taken to be equal for all hadrons. With the 
parameters $m_M=8$ MeV, $m_B=800.5$ MeV, and 
\begin{eqnarray}
A_M& =&4.95, \quad\quad C_M=6.90\cdot10^{-9}~{\rm MeV}^{1-A_M}{\rm fm}^{-3}\,, \nonumber \\
A_B & = & 6.087, \quad C_B=2.564\cdot10^{-9}~{\rm MeV}^{1-A_B}{\rm fm}^{-3}\,,\quad 
%
\end{eqnarray}
such a model not only represents the mass-integrated spectrum of all hadrons, but also it  
rather accurately reproduces the chemical FO densities of mesons $\rho_M$ and baryons $
\rho_B$ and the ratios $s/\rho_B$ and $s/\rho_M$ for chemical FO temperatures 
below 155 MeV \cite{KABPoS12}. The parameters of the center of the shock 
adiabat were fixed  as: $p_0=0$, $\rho_0=0.159\,\, {\rm fm}^{-3}$
and $\varepsilon_0=126.5~{\rm MeV~fm}^{-3}$.

The  QGP EOS is motivated by the MIT-Bag model \cite{MITBag}
\begin{eqnarray}\label{EqVII}
p_Q =A_0T^4+A_2T^2\mu^2+A_4\mu^4-B  \,, 
\end{eqnarray}
where the constants  $A_0 \simeq 2.53 \cdot 10^{-5}~{\rm MeV}^{-3}{\rm fm}^{-3}$, $A_2 
\simeq 1.51 \cdot 10^{-6} ~{\rm MeV}^{-3}{\rm fm}^{-3}$, $A_4 \simeq 1.001 \cdot 10^{-9}~{\rm 
MeV}^{-3}{\rm fm}^{-3}$, and $B \simeq 9488~{\rm MeV}~{\rm fm}^{-3}$ were found by fitting  
the  $s/\rho_B$ chemical FO data for $E_{lab} <  50$ GeV with $s/\rho_B$ values along the 
RHT adiabat and by keeping the pseudocritical temperature value 
at zero baryonic density close to 150 MeV, in agreement with lattice-QCD data \cite{Aoki}. 

Note that the above values of the coefficients $A_0, A_2$, and $A_4$ 
differ from the values  $A_0^{L}, A_2^{L}$, and $A_4^{L}$
obtained within lattice QCD \cite{Aoki} at vanishing baryonic chemical potential, but 
this difference can be attributed to the $T$ and $\mu_B$ dependence of the bag pressure 
\begin{eqnarray}
B_{eff}(T, \mu_B) &=& B - (A_0-A_0^{L} )T^4 - (A_2-A_2^{L})T^2\mu^2 \nonumber \\
& - &
(A_4 -A_4^{L})\mu^4 \,,
\end{eqnarray}
which identically generates the  QGP pressure (\ref{EqVII}) 
$p_Q =A_0^LT^4+A_2^LT^2\mu^2+A_4^L\mu^4-B_{eff} $, but with coefficients 
$A_0^{L}, A_2^{L}$, and $A_4^{L}$. The obtained result for 
$B_{eff}(T, \mu_B)$ is in line 
with the requirements of the finite-width model \cite{FWM:1,FWM:2} of quark-gluon bags. 
 
 \begin{figure}[t]
\centerline{\hspace*{-4.4mm}\includegraphics[width=84.mm]{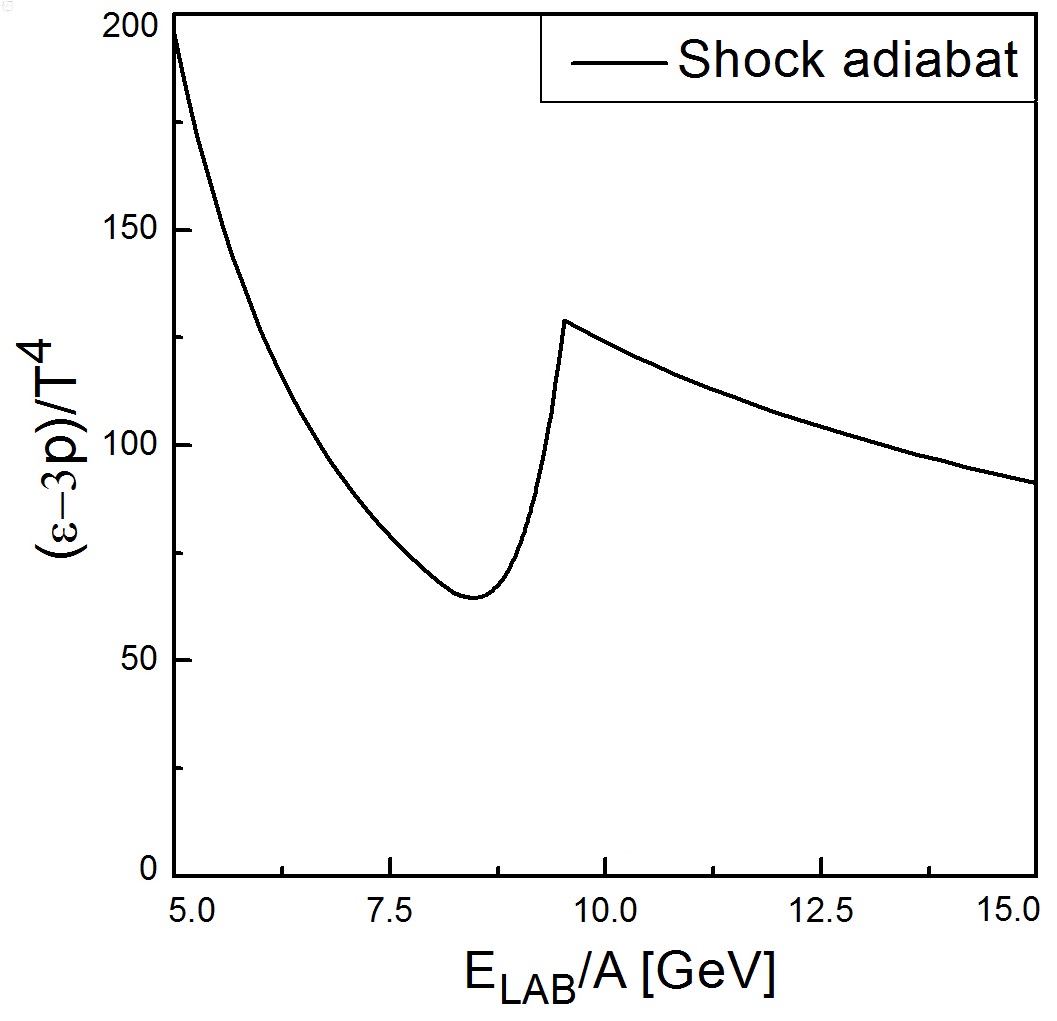}}

\vspace*{2.2mm}

\centerline{\hspace*{-2.2mm}\includegraphics[width=80.mm]{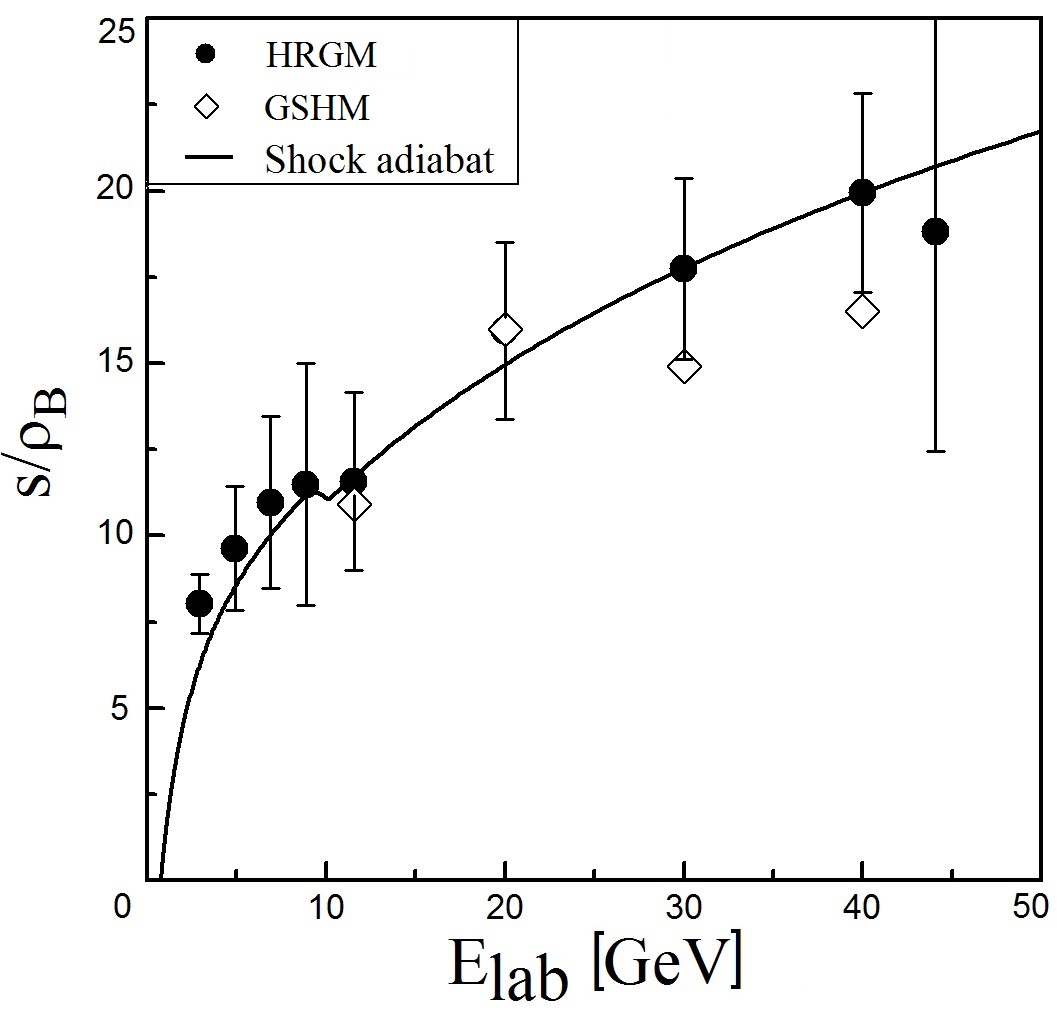}}
 \caption{
 Upper panel: 
 Energy dependence of the trace anomaly $\delta$ along the RHT adiabat of 
 the employed EOS.
 The shown $\delta$-peak generates the chemical FO $\delta$-peak  
 depicted in Fig.\ \ref{fig5} (see text for details).
 Lower panel:  Energy dependence of the entropy per baryon within the 
 HRGM (circles) and the GSHM \cite{RafLet} (diamonds).
 The solid curve represents the entropy per baryon along the RHT adiabat 
 shown in Fig.\ \ref{fig11}.
}
  \label{fig12}
\end{figure}

Using the above EOS we calculated the phase diagram and constructed the 
RHT adiabat inside all phases. 
As usual, the phase transition was found from the Gibbs criterion, 
$p_H(T,\mu_B) = p_Q(T,\mu_B)$.
The resulting RHT adiabat describes the $s/\rho_B$  chemical FO data 
well (see Fig.\ \ref{fig11}). 
The most remarkable finding is the appearance of a peak in the trace anomaly,
which exists exactly at the boundary of the mixed phase and 
the QGP (see upper panel of Fig.\  \ref{fig12}).  Comparing the 
trace-anomaly peak at chemical FO (see Fig.\ \ref{fig5}) and that on 
the RHT adiabat shown in Fig.\ \ref{fig12}, one can conclude that the
respective collision energies are very close to each other.  
This is not a coincidence, but a natural one-to-one correspondence. A slight 
shift of the position of the peaks relative to each other is due to the fact that 
the chemical FO points are discrete while the trace-anomaly peak on the RHT adiabat is 
located just between two neighboring values of collision energies $E_{lab} = 8.6$ GeV
($\sqrt{s_{NN}}=4.3$ GeV)
and $E_{lab} = 11.6$ GeV ($\sqrt{s_{NN}}=4.9$ GeV). Moreover, from the fact that inside 
the mixed phase and inside the QGP near its border with the mixed phase the quantitative 
differences between the shock, Poisson, and generalized shock adiabats 
are very small (compare, for instance, the Poisson and shock adiabats 
passing through the Chapman-Jouguet  point $A_2$ in Fig.\ \ref{fig10}) one can safely 
conclude that the $\delta$-peak on the shock adiabat and 
on the generalized one (which consists of the Poisson 
and the shock adiabat) should be located at practically the same collision 
energy. Our numerical results support 
this conclusion, but the analysis of the trace anomaly along the shock adiabat 
is much simpler than the one for the generalized shock
adiabat, which we omit here. 

The large difference in the values of trace anomaly at chemical FO and at the stage of matter 
formation can be explained by the complicated behavior of the 
trace anomaly along the Poisson adiabat, but its detailed analysis is out of the scope of  
the present work and it will be published elsewhere. Here we would like to 
stress that the found relation between the trace-anomaly peaks tells one that its 
peak seen at chemical FO is, probably, the most direct and physically 
most intuitive signal for QGP formation.

At the same time there always remains the question whether the results obtained
above are model independent. In particular, it is rather important to understand whether the 
plateaus found in the entropy per baryon can be obtained by other models of 
chemical FO. It is clear that the situation with the total number of charged pions per baryon 
is much simpler because the total number of charged pions is measured in experiment. 
Hence, only the model which correctly reproduces such an experimentally measured ratio  
should be considered to be valid. A similar situation exists
for the number of thermal pions per baryon because all viable  
thermal models use more or less the same data on the properties  of hadrons and hadronic 
resonances and, therefore, if two models correctly reproduce the total number of charged 
pions per baryon, then it is reasonable to expect that the number of thermal pions per baryon 
in these models will be also the same. 

In order to demonstrate the model independence of the
high-energy plateau in the entropy per baryon found here (see Fig.\ \ref{fig11}), we use  
an entirely different statistical model of chemical FO 
known as the generalized statistical hadronization model (GSHM) \cite{RafLet}. 
A word of caution should be added regarding the GSHM  framework,
because the chemical FO volumes found within the GSHM are in most cases essentially 
larger than the kinetic FO volumes determined by Hanbury-Brown and Twiss (HBT) 
measurements (for a detailed comparison of chemical and kinetic FO volumes see 
Ref.\ \cite{KABAndronic:09}, while for a review on HBT measurements of kinetic 
FO volumes see Ref.\ \cite{HBT}). In other words, within the GSHM the chemical FO occurs 
too late, namely only after kinetic FO, when hadrons cease to
scatter elastically and the momentum spectra of hadrons no longer changes,
while physically it should occur prior to kinetic FO.
Furthermore, very recently two successful models \cite{KAB:gs,Gupta} were suggested,
which treat the chemical FO of strange and non-strange hadrons separately. 
These models not only allow one to extremely well reproduce hadronic multiplicities 
measured in A+A collisions without any deviation of  strange charge from chemical 
equilibrium, but they also clearly demonstrate that the apparent chemical non-equilibrium 
introduced long ago in Ref.\ \cite{Rafelsky:gamma} simply reflects the fact that the strange 
hadrons experience chemical FO at a different hypersurface than the non-strange ones.  

Keeping these cautionary remarks in mind, we now compare the GSHM results for 
the entropy per baryon at chemical FO with 
the ones from our model.  As one can see from the lower panel of Fig.\ 
\ref{fig12}, despite the principal  differences between the 
HRGM \cite{KAB:13} and the GSHM \cite{RafLet}, the extracted 
values of the entropy per baryon are very close to each other. Unfortunately, the authors of  
Ref.\ \cite{RafLet} did not publish their error bars, but it is clear that the error 
bars of the GSHM should be similar to the error bars of 
the HRGM. In this case, as one can see from the lower panel of  
Fig.\ \ref{fig12}, these two different models yield, within error bars, very similar 
values for the entropy per baryon. 

 \begin{figure}[t]
\centerline{~~\includegraphics[width=88.mm]{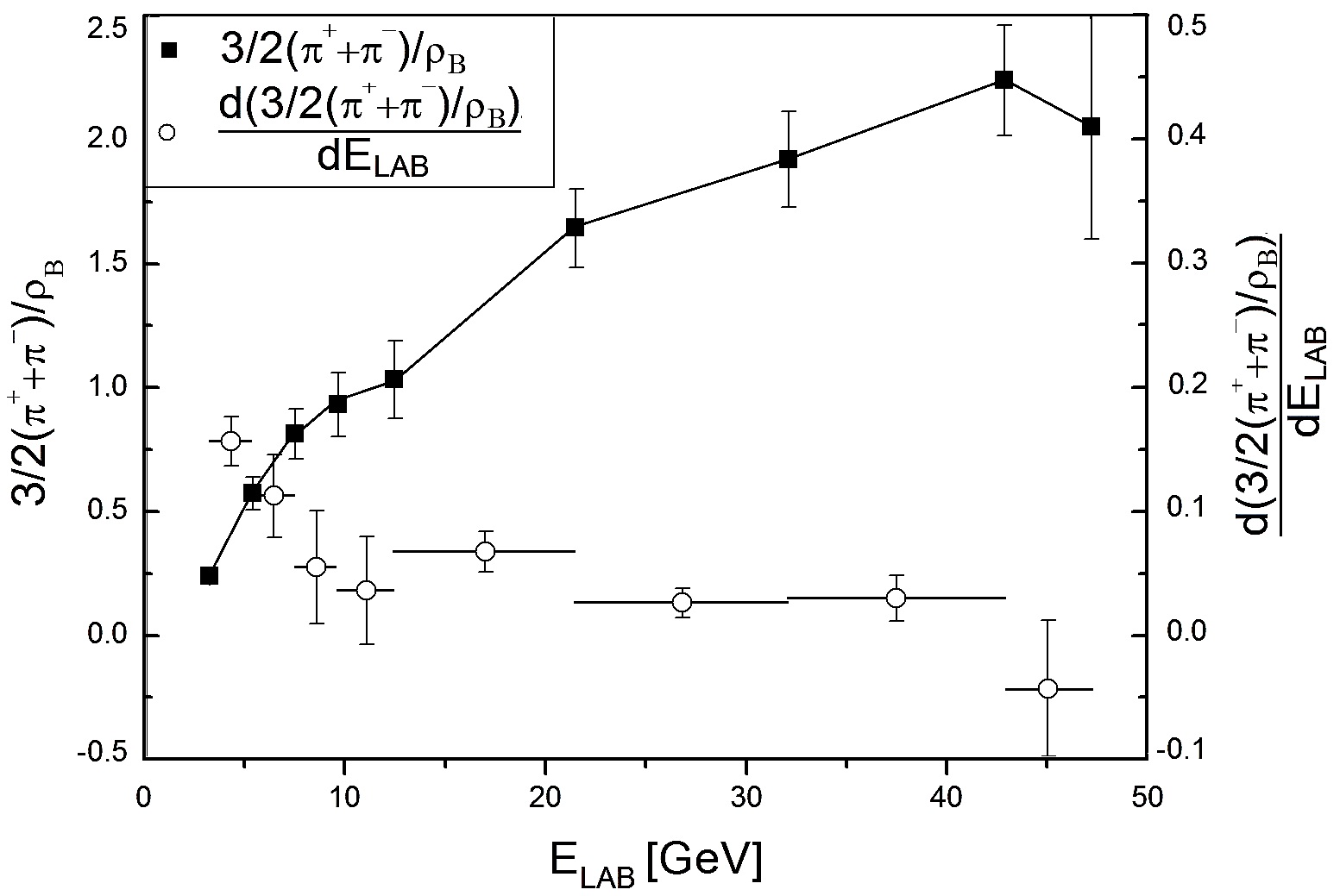}~~~}
 \caption{Energy dependence of the total number of  pions per  baryon (squares) 
 and its derivative with respect to $E_{lab}$ (horizontal bars with circles). 
 The local minimum at $E_{lab} \simeq 10$ GeV  
 in the derivative of the total pion number per baryon corresponds to the lower set of 
 plateaus found in Fig.\ \ref{fig11}. The straight lines connecting points are shown to guide
 the eye.
 }
  \label{fig13}
\end{figure}

 \begin{figure}[t]
\centerline{~~\includegraphics[width=110.mm]{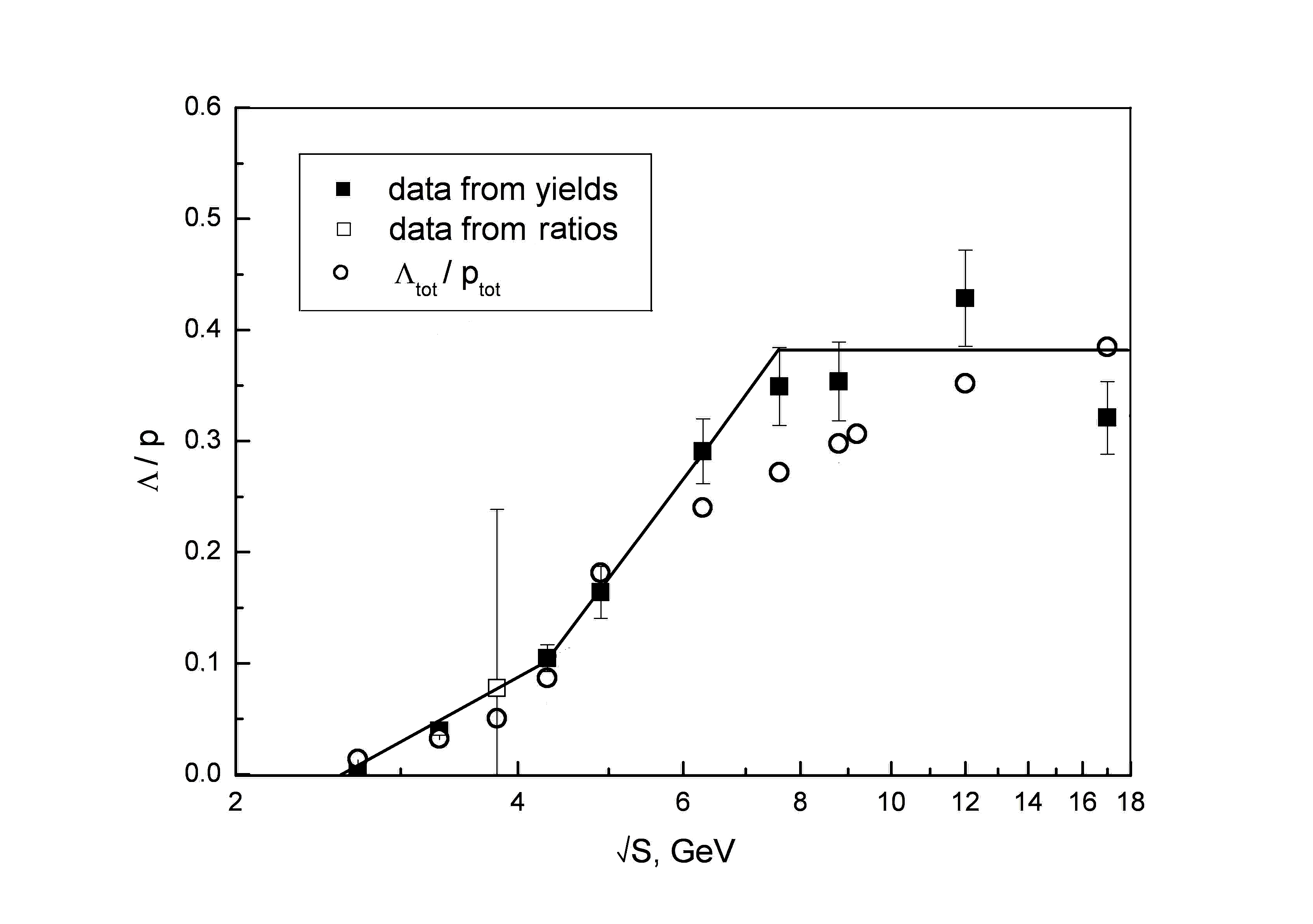}~~}
 \caption{Energy dependence of the experimental number of $\Lambda$ hyperons  
 to the experimental number of protons (squares) and their description within the
 HRGM (circles) \cite{Sagun14} which is the most recent version of 
 HRGM  \cite{KAB:gs} used here. The lines are shown to guide the eye. 
 }
  \label{fig14}
\end{figure}

One may wonder why, despite the differences between the GSHM and the HRGM,
the values for the entropy per baryon found within these models are the same
within error bars. The point is that the GSHM, like the HRGM, correctly reproduces
the multiplicity ratios of many hadrons, including pions and  baryons.
Since at low collision energies and, hence, at low freeze-out
temperatures, the entropy is simply proportional to the thermal
number of pions, one should not be surprised by the agreement of the entropy
per baryon between these two models. Therefore, it
seems that the GSHM, despite its marked differences to our version of the
HRGM, indeed also produces a plateau in the entropy per baryon at 
$E_{lab}= 20-40$ GeV. Unfortunately, the GSHM \cite{RafLet} does not describe the data at
laboratory energies below 11 GeV.

An immediate conclusion from the comparison of the entropy 
per baryon for the HRGM \cite{KAB:13} and the GSHM \cite{RafLet} is that, if 
a statistical model with or without chemical equilibrium very well reproduces all hadronic 
ratios, then one can safely extract from it both the entropy per baryon and the number of 
thermal pions per baryon. Using these quantities, one can also verify 
the existence of plateaus in their energy dependence with high confidence. 
We hope that more precise experimental data will help us to maximize this confidence in 
the future.

In order to give further indications for the plateaus discussed above,
in the following we discuss additional 
irregularities which can be obtained directly from the data. {The first is} the
behavior of the derivative of the total number of charged  
pions per baryon, or more precisly,  
the quantity  $\frac{3}{2}(\rho_{\pi^+}+ \rho_{\pi^-})/\rho_B$, with respect to 
laboratory energy, which is shown in Fig.\ \ref{fig13}.
{In order to} find it we used cubic splines and then calculated the 
corresponding derivative. 
In Fig.\ \ref{fig13} the horizontal bars correspond to the values of the 
derivative found in the middle between two neighboring energy points. 
The local minimum at $E_{lab} \simeq 10$ GeV 
in the derivative of the total pion number per baryon corresponds to
the lower sets of plateaus found in Fig.\ \ref{fig11}. One could wonder 
why for large error bars of this ratio its derivative, seemingly, has smaller errors.
The procedure of calculating the error bars shown in Fig.\ \ref{fig13} 
is given in the Appendix. 
 
{The second irregularity appears in the experimental} number of 
$\Lambda$ hyperons per proton, as shown in Fig.\ \ref{fig14}. It is necessary to say 
few words about the experimental data of hadron yields shown in Fig.\ \ref{fig14}.
For the AGS collision energies $\sqrt{s_{NN}}=$ 2.7, 3.3, and 4.3 GeV the yields of 
protons  and $\Lambda-$hyperons 
were taken, respectively, from Refs.\ \cite{AGS_p2,AGS_p1} and \cite{AGS_L}. 
Experimental yields measured at the highest AGS energy  $\sqrt{s_{NN}}=$
4.9 GeV for protons were taken from Ref.\  \cite{AGSpi2,AGS_p1}, while  
for $\Lambda$ they were given in Ref.\ \cite{AGS_L2}.  
The mid-rapidity yields of protons and lambdas 
measured  at the SPS energies $\sqrt{s_{NN}}=$
6.3, 7.6, 8.8, 12, and 17.3 GeV are given by the NA49 collaboration in Refs.\
\cite{KABNA49:17a,KABNA49:17b,KABNA49:17Ha,KABNA49:17Hb,KABNA49:17Hc,KABNA49:17phi}.
 
For a comparison, in Fig.\ \ref{fig14}  we also show the value found from 
other two ratios, $\Lambda/\pi^-$ and $p/\pi^-$, for  $\sqrt{s_{NN}}=3.8$ GeV.
This figure clearly demonstrates that there are three regimes in {the}
energy dependence of  {the} $\Lambda/p$
ratio: at  $\sqrt{s_{NN}}=4.3$ GeV the slope of this ratio visibly increases, while above   
$\sqrt{s_{NN}}=7.6$ GeV it approximately saturates.  
The change of slope at  $\sqrt{s_{NN}}=4.3$ GeV is
very similar to the prediction of Ref.\ \cite{Rafelski:82} that the mixed-phase formation can be 
identified by {a rapid} increase {in
the number of strange quarks per light quarks}.  
The present ratio is convenient since at low collision energies $\Lambda$ hyperons
are {generated} in
collisions of nucleons. {As} it seems, this mechanism works up to  
$\sqrt{s_{NN}}=4.3$ GeV,
while an appearance of the mixed phase should lead to {an} increase 
{in the} number of strange quarks and antiquarks 
due to {the} annihilation of light quark-antiquark {and gluon pairs}.
This is exactly seen from Fig.\ \ref{fig14}. Moreover, the present model predicting 
the mixed-phase formation at $\sqrt{s_{NN}}=4.3$ GeV naturally explains the change of   
{the} experimental $\Lambda/p$ ratio slope at this energy on the basis of
{the} idea of Ref.\ \cite{Rafelski:82}.

It is necessary to make a remark on the origin of the high-energy  
plateaus {seen in Fig.\ \ref{fig11}}. The main question is whether such  
plateaus provide evidence for {another cross\-over or phase transition}, 
with an associated instability of the shock adiabat. A close inspection of 
Figs.\ \ref{fig3}, \ref{fig4}, and \ref{fig5} reveals that at RHIC energy 
$\sqrt{s_{NN}}= 9.2$ GeV (or $E_{lab} \simeq 44$  GeV) \cite{KABstar:9.2} there may be a 
hint of irregularities similar to the ones associated with the low-energy 
plateaus. Unfortunately, their small amplitude compared to the neighboring points and 
very large error bars do, {at least at the moment,}
not allow us to safely interpret them as  
clear signals of some instability {related to a crossover or phase transition}.

{However, if we assume that these high-energies plate\-aus indeed
exist}, at present we see the following three possibilities {for their origin}. 
The first possibility is that they are, in fact, 
quasiplate\-aus which are similar to the ones that were found for
a pure QGP EOS in the case of Z-kind models \cite{KAB:90}. They are not 
associated with a {crossover or phase transition}, but may 
appear at high energy density
due to specific features of the {QGP EOS}. In this case 
only the low-energy plateaus correspond to the quark-gluon-hadron mixed phase. 
 
The second possibility is {that the 
quasiplateaus at $E_{lab}$ $\simeq 30 - 44$ GeV correspond to the same
crossover or phase transition which exists at lower energies. The initial
states of the system lie in the pure QGP phase between the lower and the higher 
sets of plateaus, but return to the phase boundary at the higher end of this
range of collision energies. 
This idea is further elaborated in Ref.\ \cite{Kizka2015}. 
Support for this interpretation comes from the PHENIX collaboration
which concluded from the acoustic scaling of the elliptic flow} 
that the critical endpoint may exist 
at center-of-mass energies $\sqrt{s_{NN}} \simeq 30-40$ GeV 
\cite{PHENIX14,PHENIX14b}.

Finally, in the third case, if both sets of plateaus {are due 
to instabilities caused by two separate phase-transition regions,} 
the traditional irregularities \cite{Kink,Horn,Step} 
may reflect the second phase transition. But even in this case it is unclear whether 
the onset of deconfinement begins at $\sqrt{s_{NN}} \simeq 7.6$ GeV 
($E_{lab} \simeq 30$ GeV) as suggested in Ref.\ \cite{Gazd_rev:10} or 
whether it can appear at $\sqrt{s_{NN}} \simeq 10$ GeV
as suggested in the work on quarkyonic matter \cite{QYON}. At the moment it is absolutely 
unclear whether the second phase transition is a deconfinement or a chiral-symmetry
restoring transition.
Moreover, it is an open question whether this is 
a second-order phase transition, a first-order one, or a 
crossover. From the current state of the art one can speculate that the local 
minimum of $X(\sqrt{s_{NN}})$ and the small peak 
of the trace anomaly $\delta$ seen at $\sqrt{s_{NN}} \simeq 9.2$ GeV could, probably, 
signal a very weak first-order phase transition or a second-order one, but the error bars of the 
quantities measured at this energy range are very large and, hence, to make 
a more definite conclusion we need about ten more data points and 
an experimental accuracy that is one order of magnitude better.

\section{Conclusions}

In this work we have presented irregularities at chemical FO elucidated via a high-quality fit of 
experimental particle ratios obtained by the advanced version of the hadron resonance gas 
model. The fit quality achieved, $\chi^2/dof \simeq 1.16$, 
lends our results high confidence. Among these irregularities we observed a 
dramatic jump of the effective number 
of degrees of freedom at  center-of-mass collision energies $\sqrt{s_{NN}} = 4.3-4.9$ GeV.
At $\sqrt{s_{NN}} = 4.9$ GeV we also found a local minimum of the generalized 
specific volume $X^{FO} (\sqrt{s_{NN}})$ and a sharp peak of the 
trace anomaly $\delta(\sqrt{s_{NN}})$. 

Furthermore, at 
chemical FO we found two sets of plateaus in the collision-energy dependence of the entropy 
per baryon, and of the total and the thermal numbers of pions per baryon, which were 
predicted a long time ago \cite{KAB:89a,KAB:90,KAB:91}.  We discussed in some detail
the generalized shock-adiabat model for low-energy central nuclear collisions 
and argued that the plateaus and the minimum of $X^{FO} 
(\sqrt{s_{NN}})$ found at low collision energies are generated by the 
RHT-adiabat instabilities existing at the boundary 
between the mixed phase and the QGP.  A numerical 
calculation of the RHT adiabat for a
realistic EOS of the hadronic phase allowed us to reproduce the
$s/\rho_B$ plateau and to fix the parameters of the QGP EOS.
Moreover, the trace anomaly $\delta$ calculated along the RHT adiabat 
exhibits a sharp peak at the boundary of QGP and mixed phase, which is l
ocated at a center-of-mass collision energy $\sqrt{s_{NN}}\simeq 4.46$ GeV. 
Our numerical analysis shows a one-to-one correspondence 
between the trace-anomaly peaks found on the RHT adiabat 
and at chemical FO. Therefore, it is quite possible that the trace-anomaly peak found at  
chemical FO in combination with other irregularities 
provide evidence of the phase transition to the QGP in nucleus-nucleus 
collisions. Such a conclusion {would also naturally explain} 
the change of {the slope of the} experimental $\Lambda/p$ ratio 
at $\sqrt{s_{NN}}=4.3$ GeV, due to {the enhanced production} 
of strange quarks and antiquarks {as} suggested in Ref.\ \cite{Rafelski:82}.

Also, at chemical FO we found 
a second set of plateaus at $E_{lab} \simeq 30-44 $ GeV, which, however, 
{might} not 
correspond to a phase transition or to the discussed instabilities.
Since the generalized shock-adiabat model is not applicable at this energy 
range we cannot make definite conclusions about the origin of this set of pla\-teaus. 
However, {a} remarkable fact is that a similar plateau in the collision-energy 
dependence of the entropy per baryon occurs in  the same energy range within
an entirely different model known as GSHM \cite{RafLet}. This example demonstrates
that any statistical model of chemical FO
which correctly reproduces the multiplicities of the most abundant hadrons 
such as pions and baryons 
should be able to generate similar values of the entropy per baryon because at chemical FO 
the main contribution to the entropy is provided by pions. 
Therefore, at chemical FO we may expect a model-independent behavior 
of the entropy per baryon as function of collision energy.
 
In order to make more definite conclusions about 
the plateaus found at laboratory energies $30-44$ GeV one needs more 
precise data measured in steps of $E_{lab}$ of about 100-200 MeV and a more  
realistic model of heavy-ion collisions in this range of collision energies. 
However, now it is already clear that, if an 
additional phase transition indeed exists at laboratory 
energies $30-44$ GeV, then both sets of plateaus found here may be considered as the first 
evidence of two phase transitions in QCD.

\vskip3mm

\noindent
{\bf Acknowledgments.} 
The authors are thankful to E.V.\ Shuryak and I. P.\ Yakimenko for important  comments. 
K.A.B., D.R.O., A.I.I.,  
V.V.S.,  and G.M.Z.\ acknowledge  support {from}
the National Academy of Sciences of Ukraine.
D.R.O.\ acknowledges funding {from the} 
Helmholtz Young Investigator Group VH-NG-822
{of} the Helmholtz Association and GSI, and   thanks HGS-HIRe for support.
Partial support by the grant NSH-932.2014.2 
is  acknowledged by  I.N.M.\ and L.M.S.

\section{Appendix}

In this Appendix we present a derivation of the errors for the derivative of the quadratic spline 
function $y(x)$, 
which approximates the experimental data points $y_k \pm \delta y_k$ given at the 
points $x_k$, with $k =1,2, 3, \ldots$. 
Here, $\delta y_k$ denotes the error of experimental data point $y_k$.
The derivation of errors for the cubic splines is very similar but technically more involved. 
We checked numerically that the errors for the data points for the charged pions per baryon 
(see Fig.\ \ref{fig13}) given by either the quadratic or the cubic splines are indistinguishable. 
The reason is the very smooth behavior of the approximated function. 

To approximate three successive points $y_{k-1}$,  $y_{k}$, and $y_{k+1}$ by a second-order
polynomial $y(x) = A\,x^2 +B\,x +C$, we determine the set of coefficients $A, B$, and $C$ for 
each interval $x \in [x_{k-1}; x_{k+1}]$ as a solution of
\begin{eqnarray}\label{EqAI}
y_{k-1} &=&  A\,x_{k-1}^2 +B\,x_{k-1} +C \,, \\
\label{EqAII}
y_{k} &=&  A\,x_{k}^2 +B\,x_{k} +C \,, \\
y_{k+1} &=&  A\,x_{k+1}^2 +B\,x_{k+1} +C \,.
\label{EqAIII}
\end{eqnarray}
Solving the system (\ref{EqAI})--(\ref{EqAIII}) for the coefficients $A$ and $B$, one has
\begin{eqnarray}\label{EqAIV}
A &=&  \left(  \frac{y_{k-1} - y_{k}}{x_{k-1} - x_{k}} 
-  \frac{y_{k-1} - y_{k+1}}{x_{k-1} - x_{k+1}} \right) \frac{1}{x_{k} - x_{k+1}} \,,   \quad  \\
B  &=& \frac{y_{k-1} - y_{k+1}}{x_{k-1} - x_{k+1}} -  A\,(x_{k-1} - x_{k+1}) \,.
\label{EqAV}
\end{eqnarray}
Using Eqs.\ (\ref{EqAIV}) and (\ref{EqAV}) one can determine the function 
$y^\prime (x_l) = 2 A\, x_l +B$ for each considered interval 
$x_l \in [x_{k-1}; x_{k+1}]$ and its error $\delta y^\prime(x_l)$ in the same interval 
via the formula \cite{Taylor82}
\begin{eqnarray}\label{EqAVI}
\delta y^\prime(x_l) &=& \left[ \left( \frac{\partial y^\prime (x_l)}{\partial  y_{k-1}} 
\, \delta y_{k-1} \right)^2 + \left( \frac{\partial y^\prime (x_l)}{\partial  y_{k}} 
\, \delta y_{k} \right)^2 + \right. \nonumber \\
&+& \left. \left( \frac{\partial y^\prime (x_l)}{\partial  y_{k+1}} \, \delta y_{k+1} 
\right)^2  \right]^\frac{1}{2} \,.
\end{eqnarray}
This expression was used to calculate the error for the
derivative of the total pion number per baryon 
with respect to laboratory energy, see 
Fig.\ \ref{fig13}. The expression (\ref{EqAVI}) greatly simplifies, if one chooses $x_l = x_k$
\begin{eqnarray}\label{EqAVII}
&&\delta y^\prime(x_k) = \frac{1}{|x_{k-1} - x_{k} ||x_{k} - x_{k+1}| |x_{k+1}- x_{k-1}|}  \nonumber \\
&\times&\left[ \left[ (x_{k} - x_{k+1})^2 \, \delta y_{k-1} \right]^2 + \left[ (x_{k+1}- x_{k-1})^2 \,
 \delta y_{k} \right]^2 + \right. \nonumber \\
&+& \left. \left[  (x_{k} - x_{k-1})^2 \, \delta y_{k+1} \right]^2  \right]^\frac{1}{2} \,.
\end{eqnarray}
Already this formula shows that, if the wide intervals $ (x_{k} - x_{k-1})$ and 
$(x_{k+1}- x_{k-1})$ contain a large error,
say $\delta y_{k+1} \gg \{ \delta y_{k-1}; \delta y_{k}\}$, then the resulting error 
of the derivative $\delta y^\prime(x_k)$ is the smaller the wider the 
intervals $ (x_{k} - x_{k-1})$ and $(x_{k+1}- x_{k-1})$ are. This qualitatively explains, 
why the errors of the derivative of the total pion number per baryon shown in Fig.\ \ref{fig13} 
sometimes strongly differ from the errors $\delta y_{k}$.



\begin{thebibliography}{99}

\bibitem{Kink}
M. Gazdzicki,
{Z. Phys. C}
{\bf 66}, {659}
(1995).


\bibitem{Horn}
%
M. Gazdzicki  and  M.I. Gorenstein, 
{Acta Phys. Polon. B}
{\bf 30},
{2705} (1999).

\bibitem{Step}
%
{M.I. Gorenstein,  M. Gazdzicki and  K.A. Bugaev},
{Phys. Lett. B}
{\bf 567},
(2003)
{175}

\bibitem{Gazd_rev:10}
%
M. Gazdzicki, M.I. Gorenstein  and P. Seyboth,
{Acta Phys. Polon. B}
{\bf 42},
{307} (2011).

\bibitem{Oliinychenko:12}
%
D.R. Oliinychenko, K.A. Bugaev and A.S. Sorin, 
{Ukr. J. Phys.}
{\bf 58}, {211}
(2013).


\bibitem{KAB:13}
%
K.A. Bugaev, D.R. Oliinychenko,   A.S. Sorin and G.M. Zinovjev, 
{Eur. Phys. J. A}
{\bf 49}, {30}
(2013).


\bibitem{KAB:gs}
K.A. Bugaev  et al.,
{Europhys. Lett.}
{\bf 104}, {22002}
(2013).

\bibitem{NICAWP}
K.A. Bugaev  et al.,
Ukr. J. Phys. {\bf 60},  181  (2015).

\bibitem{KABCleymans:93}
%
J. Cleymans and  H. Satz,
Zeit. Phys.  C {\bf 57}, 135 (1993).


\bibitem{KABCleymansFO}
%
J. Cleymans, K. Redlich, Phys. Rev. Lett. {\bf 81}, 5284 (1998).


\bibitem{PBM:99}
%
P. Braun-Munzinger, I. Heppe, J. Stachel, Phys. Lett. B {\bf  465},  15 (1999). 

\bibitem{KABpbm:02}
%
P. Braun-Munzinger, J. Cleymans, H. Oeschler, K. Redlich, 
Nucl. Phys. A  {\bf 697},  902 (2002).


\bibitem{KABAndronic:05}
%
A. Andronic, P. Braun-Munzinger  and  J. Stachel, 
{Nucl. Phys. A}
{\bf 772}, 
{167} (2006)
and references therein.

\bibitem{KABAndronic:09}
%
A. Andronic, P. Braun-Munzinger and J. Stachel,
Phys. Lett. B {\bf 673}, 142 (2009)
 and references therein.


\bibitem{KAB:89a}
%
K.A. Bugaev, M.I. Gorenstein, B. K\"ampfer  and V. I. Zhda\-nov,
{Phys. Rev. D}
{\bf 40},
{2903} (1989).


\bibitem{KAB:90}
%
K.A. Bugaev, M.I. Gorenstein and D.H. Rischke,
{JETP Lett.}
{\bf 52}, 
{1121} (1990).

\bibitem{KAB:91}
%
K.A. Bugaev, M.I. Gorenstein and D.H. Rischke,
{Phys. Lett. B}
{\bf 255},
{18} (1991) and references therein.
 
 \bibitem{RafLet}
 J. Letessier and J. Rafelski,
 Eur. Phys. J. A {\bf 35} (2008) 221.
 
 \bibitem{THERMUS}
%
S.~Wheaton, J.~Cleymans and M.~Hauer,  
Comput. Phys. Commun. {\bf 180}, 84 (2009).

 
\bibitem{AGS_pi1}  
%
J.L.~Klay {\it et al.,}
Phys. Rev. C {\bf 68}, 054905  (2003).  
 
\bibitem{AGSpi2}
%
L.~Ahle  {\it et al.,}
Phys. Lett. B {\bf 476}, 1 (2000).
 
\bibitem{AGS_p1}  
%
B.B.~Back {\it et al.,}
Phys.  Rev. Lett. {\bf 86}, 1970 (2001).
 
\bibitem{AGS_p2}
%
J.L.~Klay {\it et al.,}
Phys. Rev. Lett. {\bf 88}, 102301 (2002).  
 
\bibitem{AGS_L}
%
 C.~Pinkenburg {\it et al.,}
Nucl. Phys. A {\bf 698}, 495c (2002).

\bibitem{AGS_L2}
%
  S.~Albergo, R.~Bellwied, M.~Bennett, D.~Boemi, B.~Bonner, H.~Caines, W.~Christie and S.~Costa {\it et al.},
  Phys.\ Rev.\ Lett.\  {\bf 88}, 062301 (2002).
 
\bibitem{AGS_Kas}
%
P.~Chung {\it et al.,}
Phys. Rev. Lett. {\bf 91}, 202301 (2003).
 
\bibitem{KABNA49:17a}
%
S.V.~Afanasiev {\it et al.,}
Phys. Rev. C {\bf 66}, 054902 (2002).
 
\bibitem{KABNA49:17b}
%
S.V.~Afanasiev {\it et al.,}
Phys. Rev. C {\bf 69}, 024902 (2004).   
 
 
\bibitem{KABNA49:17Ha}
%
T.~Anticic {\it et al.,}
Phys. Rev. Lett. {\bf 93}, 022302 (2004).
 
\bibitem{KABNA49:17Hb}
%
S.V.~Afanasiev {\it et al.,}
Phys. Lett. B  {\bf 538}, 275 (2002).
 
\bibitem{KABNA49:17Hc}
%
C.~Alt {\it et al.,}
Phys. Rev. Lett. {\bf 94}, 192301  (2005).
 
\bibitem{KABNA49:17phi}
%
S.V.~Afanasiev {\it et al.,}
Phys. Lett. B  {\bf 491}, 59 (2000).
 
\bibitem{KABstar:9.2}
B.~Abelev {\it et al.,}  
Phys. Rev. C {\bf 81}, 024911 (2010).
 
\bibitem{KABstar:62a}
B.~Abelev {\it et al.,}  
Phys. Rev. C {\bf 79}, 034909 (2009).
 
 
\bibitem{KABstar:130a}
%
J.~Adams {\it et al.,}
Phys. Rev. Lett. {\bf 92}, 182301 (2004).
 
\bibitem{KABstar:130b}
%
J.~Adams {\it et al.,}
Phys. Lett. B {\bf 567}, 167 (2003).
 
\bibitem{KABstar:130c}
%
C.~Adler {\it et al.,}
Phys. Rev. C {\bf 65}, 041901(R) (2002).
 
\bibitem{KABstar:200a}
%
J.~Adams {\it et al.,}
Phys. Rev. Lett. {\bf 92}, 112301 (2004).
 
\bibitem{KABstar:200b}
%
J.~Adams {\it et al.,}
Phys. Lett. B {\bf 612}, (2005) 181.
 
\bibitem{KABstar:200c}
%
A.~Billmeier {\it et al.,}
J. Phys. G {\bf 30}, S363 (2004).

\bibitem{KAB:89}
%
K.A. Bugaev  and  M.I. Gorenstein,
{Z. Phys. C}
{\bf 43}, 
{261}  (1989) and references therein.
  
 \bibitem{KAB:89b}
 %
K.A. Bugaev, M.I. Gorenstein and V.I. Zhda\-nov,  
{Teor. Mat. Fizika (in Russian)}
{\bf 80},
{138} (1989).

  \bibitem{lQCD}
{Sz. Borsanyi et al.,}
{JHEP}
{\bf  1208}
(2012)
{053}.

\bibitem{Mish:78}
%
V.M. Galitskij  and I.N. Mishustin,
{Phys. Lett. B}
{\bf 72},
{285} (1978).


\bibitem{Horst:80}
H. St\"ocker,  G. Graebner, J.A. Maruhn  and W. Greiner,
{Phys. Lett. B}
{\bf 95},
{192} (1980).

\bibitem{Kamp:83}
B. K\"ampfer,
{J.  Phys. G}
{\bf 9},
{1487} (1983).

\bibitem{Horst_PR:86}
H. St\"ocker   and W. Greiner,
{Phys. Rep.}
{\bf 137},
{277} (1986) and references therein.

\bibitem{Barz:85}
H.W Barz., L.P. Csernai,  B. K\"ampfer  and  B. Lukacs,
{Phys. Rev. D}
{\bf 32},
{115} (1985).



\bibitem{Landavshitz}
L.D. Landau  and E.M. Lifshitz,
{\it Fluid Mechanics}, 
Pergamon, New York, (1979).
{}

\bibitem{Zeldovich}
Y.B. Zel'dovich  and Y.P. Raiser,
{\it Physics of shock waves and high 
temperature hydrodynamic phenomena},
Academic, New York, (1967).
{}


\bibitem{Rozhd}
B.L. Rozhdestvensky   and N.N. Yanenko,
{\it Systems of Quasi-Linear Equations},
Nauka, Moscow, (1978).




\bibitem{Satarov:11}
A.V. Merdeev, L.M. Satarov  and I.N. Mishustin,
{Phys. Rev. C}
{\bf 84},
{014907} (2011).

\bibitem{3fluid}
%
Yu.B. Ivanov, V.N. Russkikh and V.D. Toneev, 
Phys. Rev. C {\bf 73}, 044904 (2006). 


\bibitem{Walecka}
J.D. Walecka,
{Ann. Phys.}
{\bf 83},
{491} (1974).
 

\bibitem{Zimanyi}
J. Zimanyi  et al.,
{Nucl. Phys. A}
{\bf 484},
{647} (1988).

\bibitem{Bethe:42}
{H.A. Bethe,}
{Office of Scientific Research and Development Report No.}
{\bf 545},
{25} (1942).

\bibitem{KSH:96} 
%
K.A. Bugaev,  
Nucl. Phys. A {\bf 606}, 559 (1996).

\bibitem{KSH:99} 
K.A. Bugaev, M.I. Gorenstein and W. Greiner,
J. Phys. G {\bf 25}, 2147 (1999).

\bibitem{MITBag}
A. Chodos   et. al.,
{ Phys. Rev. D}
{\bf 9},
{3471} (1974).

\bibitem{KABPoS12}
K.A. Bugaev et al.,
{PoS Baldin ISHEPP XXI}
{\bf 2012},
{017} (2012).


\bibitem{Aoki}
Y. Aoki  et al.,
{Phys. Lett. B}
{\bf 643},
{46} (2006).

\bibitem{FWM:1}
K. A. Bugaev, V. K.  Petrov and  G. M. Zinovjev,
{Europhys. Lett.} {\bf 85},  22002 (2009).

\bibitem{FWM:2}
K. A. Bugaev, V. K.  Petrov and  G. M. Zinovjev,
{Phys. Rev.}  C.  {\bf 79}, 054913 (2009).

\bibitem{HBT}
%
M. Lisa, S. Pratt, R. Soltz, U. Wiedemann, Ann. Rev. Nucl. Part. Sci. 
{\bf 55} (2005) 357.

\bibitem{Gupta} 
%
S. Chatterjee,  R. M. Godbole   and   S. Gupta,
Phys.  Lett.  B {\bf 727}, 554 (2013).

\bibitem{Sagun14}
%
V. V. Sagun, 
Ukr. J. Phys.  {\bf 59},  755 (2014).


\bibitem{Rafelsky:gamma}
%
J. Rafelski, Phys. Lett. B {\bf 62}, 333 (1991).  

\bibitem{Rafelski:82}
%
J. Rafelski and B. M\"uller, Phys. Rev. Lett. {\bf 48}, 1066 (1982).

\bibitem{Kizka2015}
%
 V.  A. Kizka,  V. S. Trubnikov,    K. A. Bugaev and D. R. Oliinychenko,
 arXiv:1504.06483   [hep-ph]   (2015).


\bibitem{PHENIX14}
%
R. Lacey et. al, Phys. Rev. Lett. {\bf 112}, 082302 (2014).  

\bibitem{PHENIX14b}
%
R.  A. Lacey, Talk  at  24-th International Conference on Ultra-Relativistic Nucleus-Nucleus Collisions ``Quark Matter 2014", 19-24 May,  2014, Darmstadt, Germany; 
 arXiv:1408.1343 [nucl-ex].
 

\bibitem{QYON}
A. Andronic  et. al,  
Nucl. Phys.  A {\bf 837},  65 (2010).

\bibitem{Taylor82}
%
J. R. Taylor, {\it ``An introduction to error analysis'', University Science Book Mill Valley, California (1982).}


\end{thebibliography}
\end{document}